\documentclass[aps, twocolumn, nofootinbib, superscriptaddress, floatfix]{revtex4-1}
\newcommand{\myparagraph}[1]{\par\bigskip\noindent\textbf{#1}}
\usepackage[section]{placeins}
\usepackage{silence}
\usepackage{float}
\usepackage[utf8x]{inputenc}
\usepackage[T2A]{fontenc}
\usepackage{booktabs}
\usepackage{tabularx}
\usepackage[normalem]{ulem}
\usepackage{units}
\usepackage[english]{babel}
\usepackage{xspace}
\usepackage{paralist}
\usepackage{comment}
\usepackage{amsmath,amssymb,bm} 
\usepackage{graphicx}
\usepackage{color,xcolor}
\usepackage{soul}
\definecolor{antiquewhite}{rgb}{0.98, 0.92, 0.84}
\sethlcolor{antiquewhite}

\usepackage{lineno}
\usepackage{longtable}
\usepackage[]{hyperref}
\usepackage[abs]{overpic}
\newcommand{\pn}{\ensuremath{p\leftrightarrow n}\xspace}
\newcommand{\bbn}{\textsc{bbn}}

\newcommand{\decay}{{\text{\rm decay}}}
\newcommand{\conv}{\text{conv}}

\newcommand{\He}{\ensuremath{^{4}\text{He}}\xspace}
\newcommand{\Br}{\text{Br}}

\newcommand{\eff}{\text{eff}}
\newcommand{\dec}{\text{dec}}

\newcommand{\EM}{\textsc{em}}

\newcommand{\gev}{\text{ GeV}}
\newcommand{\s}{\text{ s}}
\newcommand{\mn}{m_{N}}
\newcommand{\tn}{\tau_{N}}

\renewcommand\thesection{\arabic{section}}

\begin{document}
\title{Improved BBN constraints on heavy neutral leptons}

\author{A.~Boyarsky}
\affiliation{Instituut-Lorentz for Theoretical Physics, Universiteit Leiden, Niels Bohrweg 2, 2333 CA Leiden, The Netherlands}

\author{M.~Ovchynnikov}
\affiliation{Instituut-Lorentz for Theoretical Physics, Universiteit Leiden, Niels Bohrweg 2, 2333 CA Leiden, The Netherlands}

\author{O.~Ruchayskiy}
\affiliation{Niels Bohr Institute, University of Copenhagen, Blegdamsvej 17, DK-2100 Copenhagen, Denmark}
\author{V.~Syvolap}
\affiliation{Niels Bohr Institute, University of Copenhagen, Blegdamsvej 17, DK-2100 Copenhagen, Denmark}

\begin{abstract}
We constrain the lifetime of thermally produced Heavy Neutral Leptons (HNLs) from Big Bang Nucleosynthesis. We show that even a small fraction of mesons present in the primeval plasma leads to the over-production of the primordial helium-4. This constrains the lifetime of HNLs to be $\tau_{N}<0.02$~sec for masses above the mass of pion (as compared to 0.1 sec reported previously). In combination with accelerator searches, this allows us to put a new lower bound on the HNLs masses and to define the ``bottom line'' for HNL searches at the future Intensity Frontier experiments.
\end{abstract}

\maketitle

\myparagraph{Introduction.} Heavy neutral leptons (HNLs or right-handed neutrinos) are hypothetical particles capable of explaining neutrino masses and oscillations~\cite{Alekhin:2015byh} and resolving other beyond-the-Standard-Model phenomena: the origin of the baryon asymmetry of the Universe~\cite[see e.g.][]{Canetti:2012zc} and the nature of dark matter~\cite{Boyarsky:2018tvu}.
Extending the Standard Model with exactly three HNLs with masses below the electroweak scale and not other heavy physics allows explaining all three beyond-the-Standard-Model phenomena~\cite{Asaka:2005an,Asaka:2005pn,Boyarsky:2009ix}.
An attractive feature of the models with MeV--GeV scale HNLs is that they can be tested by a combination of accelerator searches 
at the LHC~(\cite[see e.g.][]{Alekhin:2015byh,Boiarska:2019jcw} and refs.\ therein) and future intensity frontier experiments~\cite[see e.g.][]{Beacham:2019nyx}.

The single requirement that HNLs contribute sizeably to the masses of neutrinos combined with existing accelerator exclusions does not allow putting a lower bound on the HNL mass. \emph{Such a lower bound can be obtained if we add to the picture constraints from Big Bang nucleosynthesis (BBN)}. Primordial abundances of \He and D are measured with high accuracy~\cite{Aghanim:2018eyx,Izotov:2014fga,Aver:2015iza,Peimbert:2016bdg,Fernandez:2018xx,Valerdi:2019beb}.
Their agreement with the Standard Model predictions~\cite{Pitrou:2018cgg,Arbey:2011nf,Lisi:1999ng,Mendoza:1999ki,Pisanti:2007hk} serves as one of the ``pillars'' of modern cosmology.

The success of the Standard Model-based BBN (SBBN) predictions permits using it to constrain hypothetical particles with lifetimes as small as $\mathcal{O}(10^{-2})\s$ (\cite[see e.g.][]{Pospelov:2010hj} for review). In particular, decays of HNLs in MeV-temperature plasma affect two observable quantities: (i) the abundances of light elements (in this paper we consider only \He); (ii) the effective number of relativistic species, $N_\eff$. This was used to put an upper bound on HNL lifetime~\cite{Dolgov:2000pj,Dolgov:2000jw,Dolgov:2003sg,Fuller:2011qy,Ruchayskiy:2012si,Hernandez:2013lza,Hernandez:2014fha,Vincent:2014rja,Gelmini:2019wfp,Kirilova:2019dlk,Gelmini:2020ekg,Sabti:2020yrt}. These bounds were obtained using decays of HNLs into electromagnetic particles and neutrinos. 

\emph{The goal of this paper is to derive BBN bounds for HNLs that can decay into mesons.}
The effect of mesons on the MeV plasma is qualitatively different as they interact with protons and neutrons via strong interactions. Although lifetimes of mesons, $\tau_{\text{meson}}\sim 10^{-8}\s$, are orders of magnitude smaller than any relevant BBN time scales, they can be present in the plasma as long as HNLs are still abundant and decay. $p\to n$ and $n \to p$ reactions driven by mesons have a cross-section $\mathcal{O}(10^{16})$ larger than that of weak processes that are normally responsible for
the $\pn$ conversion in the absence of mesons (e.g. in Standard Model BBN). Moreover, these reactions, in both directions, have no threshold and roughly equal cross-sections (due to isotopic symmetry).

\begin{figure*}
    \centering
    \includegraphics[width=0.5\textwidth]{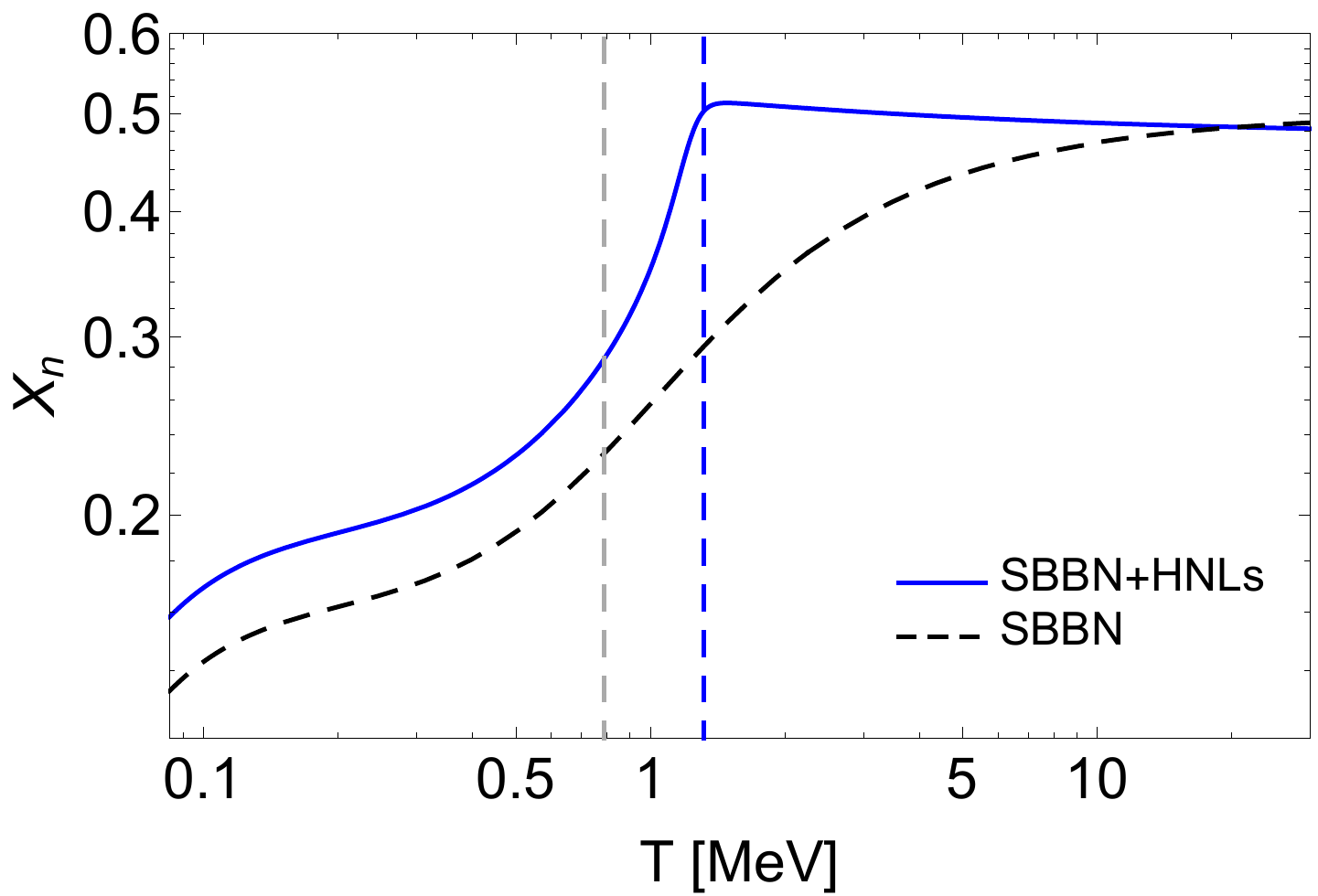}~\includegraphics[width=0.5\textwidth]{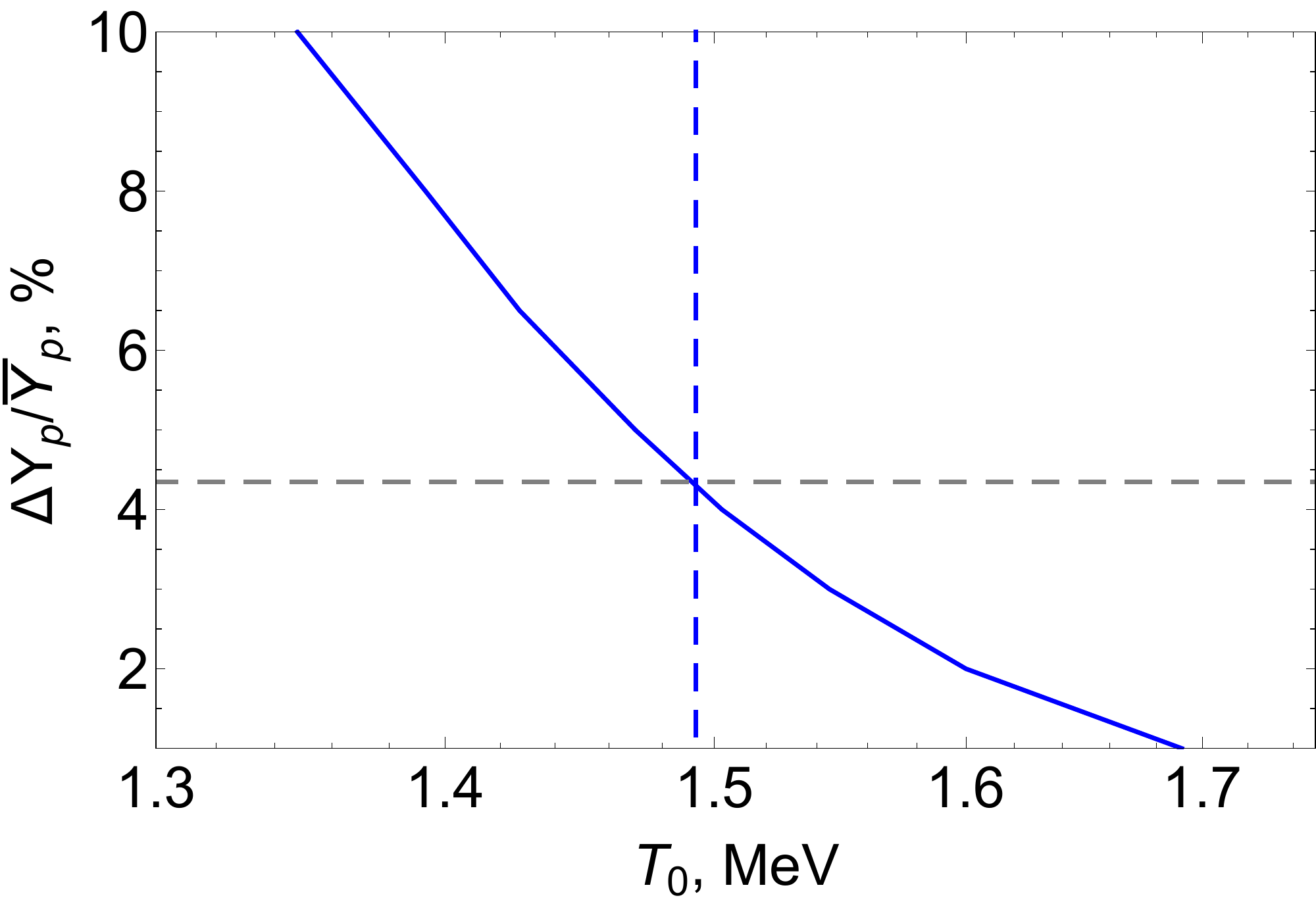}
    \caption{\textit{Left panel}: temperature evolution of the neutron abundance  $X_{n} = n_{n}/(n_{n}+n_{p})$ in the presence of pions from decays of an HNL with mass $m_{N} = 400$~MeV and lifetime $\tau_{N}=0.03\s$. Below $T\simeq 100\text{ MeV}$, pions drive the neutron abundance  to $X_{n} \approx 0.5$. At temperatures $T_{0}\simeq 1.3\text{ MeV}$ (the blue vertical dashed line) pions disappear, and $X_{n}$ starts relaxing towards its SBBN value but does not reach it.
    After the neutron decoupling (the gray vertical line) $X_{n}$ evolves mainly due to the neutron decays.
    \textit{Right panel}: a relation between the temperature $T_{0}$ (defined by Eq.~\eqref{eq:mesons-constraint-temp}) and corrections to the \He abundance, as compared to the SBBN central value $\bar{Y}_{p}\approx 0.247$. It corresponds to the case of when only charged pions are present in plasma. The gray horizontal line corresponds to maximally allowed correction $\Delta Y_p/\bar{Y}_p = 4.35\%$ that we adopt in this work (see Appendix~\ref{app:he-measurements-errors}). The intersection of gray and colored lines defines the temperature $T_{0}^{\text{min}}$. Its actual value in presence of all mesons always lies close to $T_{0}^{\text{min}}\approx 1.5$~MeV independently on the HNL mass (see Appendix~\ref{app:mesons-numeric}).}
    \label{fig:mesons-Xn-behavior}
\end{figure*}

\emph{Therefore, the presence of
even a small fraction of mesons in the plasma quickly equilibrates the number densities of protons and neutrons.} 
Once HNLs decay, mesons also disappear (instantaneously as compared to the time scales relevant for BBN), and the neutron-to-proton ratio $n_n/n_p$ relaxes solely due to the SM (weak) processes. If the HNL's lifetime is short enough, $n_n/n_p$ relaxes to its SM value before weak reactions freeze out, leaving no observable effect.
However, if HNLs (and mesons) survive until $T\simeq 1.5$~MeV and below, there is \emph{not enough time} to completely relax down to the SBBN value. This residual effect leads to a strong upper bound on the HNL lifetime.

In this work, we demonstrate for the first time that the meson-driven effect strengthens the upper bound on the HNL lifetime by a factor $\sim 5$ (down to $0.02\s$) as compared to the previous work~\cite{Dolgov:2000jw}.
In the context of dark scalars, the meson driven $\pn$ conversion and its influence on BBN were studied in~\cite{Pospelov:2010cw,Fradette:2017sdd,Fradette:2018hhl,Reno:1987qw,Kohri:2001jx,Kawasaki:2004qu}.

For temperatures corresponding to such short lifetimes, all Standard Model particles are in thermal equilibrium. 
This makes all other effects of HNLs on BBN irrelevant and allows deriving the bounds purely analytically, avoiding any computations of complicated Boltzmann equations.

\myparagraph{Meson-driven $\pn$ conversion.}
Sufficiently heavy HNLs can decay into mesons $h = \pi, K$, \textit{etc}
(see~\cite{Gorbunov:2007ak,Bondarenko:2018ptm} or Appendix~\ref{app:hadronic-hnl-decays}).
Charged pions drive the $\pn$ conversion via~\cite{Pospelov:2010cw}
\begin{equation}
    \pi^{-}+p \rightarrow n + \pi^{0}/\gamma, \quad \pi^{+}+n \rightarrow p+ \pi^{0}.
    \label{eq:mesons-processes}
\end{equation}
The cross-section of these reactions is very large:
\begin{equation}
\frac{\langle\sigma_{\pn}^{\pi}v\rangle}{\langle\sigma_{p\leftrightarrow n}^{\text{Weak}}v\rangle} \simeq \frac{1}{G_{F}^{2}m_{p}^{2}T^{2}} \sim 10^{16}\left(\frac{1\text{ MeV}}{T}\right)^{2},
\label{eq:StrongWeakRates}
\end{equation}
Large cross-section, absence of threshold and isotopic symmetry of these processes mean that if pions are present in the plasma in the amounts at least comparable with that of baryons, they drive the number densities of protons and neutrons to equal values, $n_n/n_p \simeq \langle \sigma_{p\to n}^{\pi} v\rangle/\langle \sigma_{n\to p}^{\pi}v\rangle\simeq 1$.\footnote{For each of the processes~\eqref{eq:mesons-processes}, there are no inverse reactions. Indeed, $\pi^{0}$ decays very fast, whereas $\gamma$s quickly lose their energy. Therefore, the conversion~\eqref{eq:mesons-processes} is highly non-equilibrium, and the corresponding value of $n_{n}/n_{p}$ is not given by the usual Boltzmann exponent.} The effect of kaons is qualitatively similar, but leads to a slightly different neutron-to-proton ratio (Appendix~\ref{app:mesons-numeric}).  

The impact of this effect on primordial \He abundance depends on how long mesons remain present in plasma in significant amounts.
Once mesons are created, they can \textit{(i)} scatter and lose energy; \textit{(ii)} decay; \textit{(iii)} participate in $p\leftrightarrow n$ conversion.
The corresponding rates are very different: at MeV temperatures and below, $\Gamma^h_{\rm scat} \gg \Gamma^h_\decay \gg \Gamma^h_{\pn}$ (see~\cite{Kohri:2001jx}).
The instantaneous number density of mesons is an interplay between their production (via decays of HNLs) and their decays:
\begin{equation}
   n_{h}^{\text{inst}} = n_{N}(T)\cdot \Br_{N\to h} \frac{\Gamma_{N,\text{dec}}}{\Gamma_{h, \text{dec}}} = n_{N}(T)\cdot \Br_{N\to h} \frac{\tau_h}{\tau_N}.
\end{equation}
Here, $\Br_{N \to h}$ is the branching of HNLs into mesons (Appendix~\ref{app:hadronic-hnl-decays}). $n_N(T)$ is the number density of HNLs. We consider here HNLs that were produced thermally and decouple at some temperature $T_{\dec}$ (Appendix~\ref{app:hnls-population}).\footnote{Besides thermal production, out-of-equilibrium production mechanisms of HNLs exist~\cite[see e.g.][]{Dodelson:1993je,Shi:1998km}, that we leave for future works (see, however,~\cite{Gelmini:2020ekg} where a part of this parameter space was explored).} Therefore,
\begin{equation}
    n_N(T) = \left(\frac{a_{\dec}}{a(T)}\right)^{3}\cdot n_{N}^{\dec}\cdot e^{-\frac{t(T)}{\tn}},
\end{equation}
where $n_{N}^{\dec}$ is the HNL number density at decoupling, and $a(T)$ ($a_\dec)$ is the scale factor at temperature $T$ (correspondingly, at HNL decoupling).

\begin{figure*}
    \centering  \includegraphics[width=0.5\textwidth]{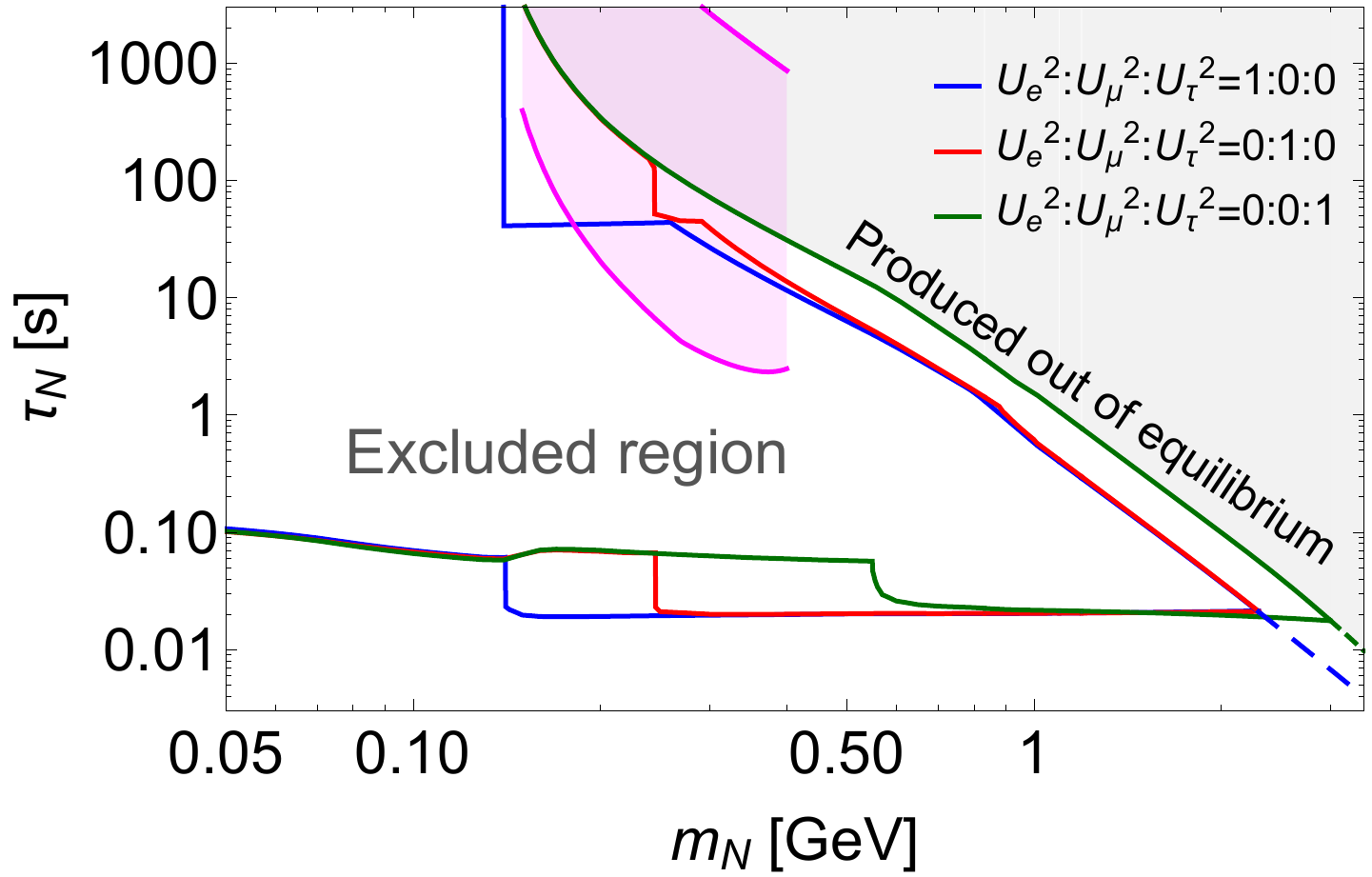}~\includegraphics[width=0.5\textwidth]{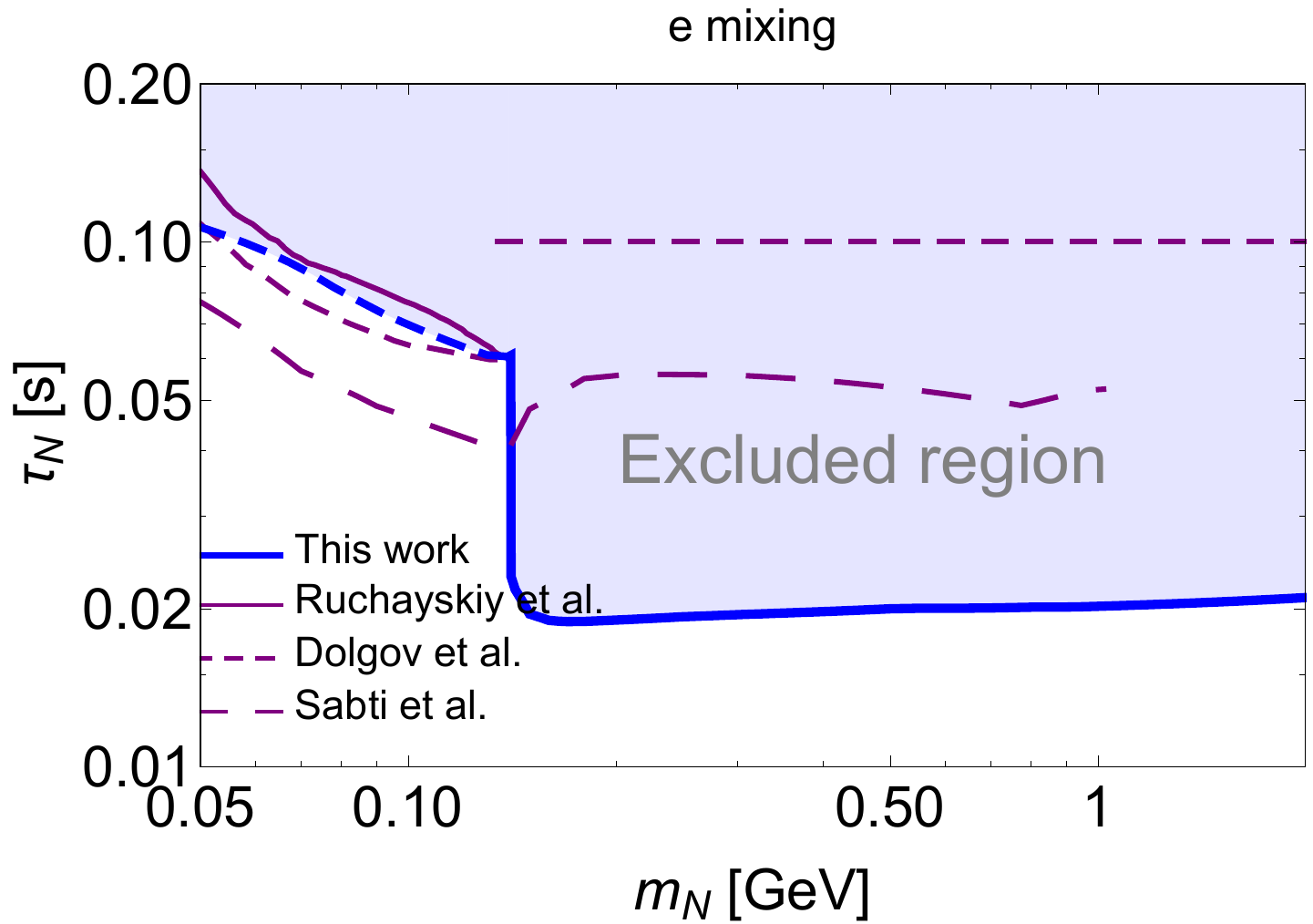}
    \caption{\textit{Left panel}: BBN bounds on HNL lifetime for different mixing patterns. The gray region is excluded as a result of this work (for masses below pion threshold we use the results of~\cite{bbn-large}). The magenta shaded region corresponds to the domain excluded in~\cite{Gelmini:2020ekg}. 
    We considered HNLs that are produced thermally and sufficiently short-lived, such that they do not survive until the onset of nuclear reactions (Appendix~\ref{app:mesons-dissociation}). \textit{Right panel}: comparison of the results of this work (thick blue line) with  the results of the previous works ~\cite{Ruchayskiy:2012si,Dolgov:2000jw,Sabti:2020yrt} (purple lines) assuming mixing with electron flavor only. Notice that other works have adopted different values for the maximally admissible \He abundance when deriving their bounds: $Y_{p,\max} = 0.2696$ in~\cite{Ruchayskiy:2012si,Dolgov:2000jw} and $Y_{p,\max} = 0.253$ in~\cite{Sabti:2020yrt} as compared to $Y_{p,\max} = 0.2573$ in this work (see text for details).}
    \label{fig:bbn-bounds-comparison} 
\end{figure*}

The number of $\pn$ reactions per nucleon occurring after time $t\gg \tn$ (or below some corresponding temperature $T(t)$) is thus
\begin{multline}
N_{\pn}^{h}(T) = \sum_h\int \limits_{t(T)}^{\infty}dt \ n_{h}^{\text{inst}}(T) \cdot \langle \sigma_{\pn}^{h} v\rangle \approx \\ \approx    \left(\frac{a_{\dec}}{a}\right)^{3}\frac{n_{N}^{\dec}}{n_{B}}\cdot e^{-\frac{t(T)}{\tn}} \cdot \Br_{N\to h}\cdot P_{\conv},
\label{eq:conversion-number}
\end{multline}
where $n_{B}$ is the baryon number density, the sum goes over meson species and $P_{\conv}$ is the probability for a single meson to interact with nucleons before decaying:
\begin{equation}
    \label{eq:Pconv}
    P_{\conv} \simeq \frac{n_{B}\cdot \langle \sigma_{p\leftrightarrow n}^{h}v\rangle}{\Gamma^{h}_{\decay}}.
\end{equation} 
At $\mathcal{O}(1\text{ MeV})$ temperatures, $P_{\conv} \sim 10^{-2}-10^{-1}$, see Appendix~\ref{app:mesons-xsection}. 

The meson driven conversion keeps the value $n_{n}/n_{p} \simeq 1$ roughly until a temperature $T_{0}$ when the number of reactions drops below one,
\begin{equation}
N_{\pn}^{h}(T_{0})\simeq 1,
\label{eq:mesons-constraint-temp}
\end{equation}
and weak SBBN reactions start to relax the $n/p$ ratio down to its SBBN value, see Fig.~\ref{fig:mesons-Xn-behavior} (left panel). However, if $T_{0}$ is close enough to the freeze-out of weak \pn processes, occurring at $T_n \simeq 0.8$~MeV, the relaxation is not complete (Fig.~\ref{fig:mesons-Xn-behavior}, right panel). This leads to a positive correction $\Delta (n_{n}/n_{p})$ as compared to the SBBN case, which translates to an increase of the \He abundance $\Delta Y_{p}$.

In this way, the \emph{upper bound} on the \He abundance $Y_{p,\text{max}}$ is translated to the \emph{lower bound} $T_{0}\ge T_{0}^{\text{min}}$. Together with the relations~\eqref{eq:conversion-number}--\eqref{eq:mesons-constraint-temp}, this allow us to find an upper limit on the HNL lifetime $\tau_{N}$:
\begin{equation}
    \tau_{N} \lesssim \frac{t(T_{0}^{\text{min}})}{\ln\left[\sum_h\left(\frac{a_{\text{dec}}}{a_{0}}\right)^3\frac{n_{N}^{\text{dec}} P_{\conv}\Br_{N\to h}}{n_{\gamma}(T_0^{\text{min}}) \eta_B}\right]}.
    \label{eq:meson-constraint-parametric}
\end{equation}
Here, $n_\gamma$ is the number density of photons, $\eta_B$ is the baryon-to-photon ratio, and $t(T)$ is time-temperature relation. $t(T)$ is given by the Standard Model relation: $t(T) = \frac{M_*}{2T^2}$, with $M_* = \frac{M_{\text{Pl}}}{1.66\sqrt{g_{*}}}$ the reduced Planck mass, where $g_{*}(T) \simeq 10.6$ for $T\simeq 1-2$~MeV.\footnote{This is indeed the case for short-lived HNLs with $\tau_{N}\ll 0.1\s$.}

Let us rewrite the logarithmic factor  in~\eqref{eq:meson-constraint-parametric} as
\begin{equation}
    \left(\frac{a_{\text{dec}}}{a_{0}}\right)^3\frac{n_{N,\dec}}{n_{\gamma}(T_0^{\text{min}})} = \frac{n_{N,\dec}}{n_{\gamma}(T_{\dec})}\cdot \left(\frac{a_{\dec}T_{\dec}}{a_{0}T_{0}^{\min}}\right)^{3}.
\end{equation}
HNLs with $\mn\gtrsim m_{\pi}$ and lifetimes $\tn\ll 0.1\s$ decouple while being ultrarelativistic, $T_{\dec}\gg m_{N}$ (see Appendix~\ref{app:hnls-population}) implying $n_{N,\dec}/n_{\gamma}(T_{\dec}) \approx 3/2$. 
In SBBN at temperatures $T\gtrsim 1$~MeV, all particles are at local equilibrium, which define the dynamics of the scale factor:
\begin{equation}
\left(\frac{a_{\dec}T_{\dec}}{a_{0}T_{0}^{\min}}\right)^3\approx \frac{g_{*}(T_{0}^{\text{min}})}{g_{*}(T_{\dec})} \simeq \frac{1}{8}.
\label{eq:a-scaling}
\end{equation}
Decays of heavy HNLs violate thermal equilibrium at $\mathcal{O}(1\text{ MeV})$ and the scaling~\eqref{eq:a-scaling} is not valid. This leads to an additional decrease of this ratio by a factor of $0.1-0.6$ for HNL masses $m_{\pi}\lesssim m_{N}\lesssim 3$~GeV (we will use $\frac{1}{3}$ for normalization below), see Appendix~\ref{app:hnls-population}.

This results in 
\begin{equation}
  \boxed{\tau_{N} \lesssim \frac{0.023\left(\frac{1.5\text{ MeV}}{T_{0}^{\text{min}}}\right)^{2} \s}{1+0.07\ln\left[\frac{P_{\conv}}{0.1}\frac{\Br_{N\to h}}{0.4} \frac{2n_{N,\dec}}{3n_{\gamma}(T_{\dec})}\cdot 24 \left(\frac{a_{\dec}T_{\dec}}{a_{0}T_{0}^{\min}}\right)^{3}\right]}.}
  \label{eq:mesons-constraint}
\end{equation}
Using values of $\Br_{N\to h}$, $P_{\text{conv}}$ and the scale factors ratio (Appendices~\ref{app:hadronic-hnl-decays}, \ref{app:mesons-xsection}, \ref{app:hnls-population} correspondingly), we conclude that the logarithm term in~\eqref{eq:mesons-constraint} is $\mathcal{O}(1)$ for HNLs in the mass range $m_{N} = \mathcal{O}(1\text{ GeV})$ and affects the overall bound very weakly. Therefore,  the bound depends only on $T_{0}^{\min}$.

The presence of mesons increases the \He abundance. Therefore, to fix $T_{0}^{\text{min}}(m_N)$, we need to adopt an upper bound on the primordial \He abundance, $Y_{p,\text{max}}$, that is consistent with measurements~\cite{Tanabashi:2018oca}. The smallest error bars come from measuring $Y_{p}$ in low-metallicity interstellar regions and  extrapolating its value to zero metallicity (pioneered in~\cite{Izotov:2013waa}). Several  groups~\cite{Izotov:2014fga,Aver:2015iza, Peimbert:2016bdg,Fernandez:2018xx,Valerdi:2019beb} have determined $Y_{p}$ using this method, albeit with  different data and assumptions. The resulting scatter between results is larger than the reported error bars. 
We treat this difference as an additional systematic uncertainty and adopt the maximal value $Y_{p,\text{max}} = 0.2573$ (see Appendix~\ref{app:he-measurements-errors}). The maximally allowed relative deviation is therefore 
\begin{equation}
\Delta Y_{p}/Y_{p,\text{\sc sbbn}} \approx 4.35\%.
\label{eq:he-correction-maximal}
\end{equation}
To relate $\Delta Y_{p}$ and $T_{0}^{\text{min}}$, we study how the $n_{n}/n_{p}$ ratio is relaxed below $T_{0}$.
The relaxation occurs solely via the SBBN reaction,
\begin{equation}
\frac{dX_{n}}{dt} = \Gamma^{\text{\sc sbbn}}_{p\to n}(1-X_{n}) - \Gamma^{\text{\sc sbbn}}_{n\to p}X_{n}, \quad X_{n} = \frac{n_{n}}{n_{n}+n_{p}},
\label{eq:Xn-equation}
\end{equation}
albeit with the altered initial condition $X_n(T_{0}) = X_n^h \simeq 1/2$. ($\Gamma^{\text{\sc sbbn}}_{\pn}(t)$ are SBBN rates, see~\cite{Pitrou:2018cgg}). 
Non-SBBN value of $X_n(T_{0})$ \emph{is the dominant  effect} of short-lived HNLs on $Y_p$. 
At temperatures $T\lesssim T_{0}$, for HNLs with lifetimes $\tau_{N}\lesssim 0.02\s$, all other quantities that are relevant for BBN dynamics -- $\eta_{B}$, time-temperature relation, the nuclear reactions chain -- remain the same as in SBBN, which is because most of HNLs are no longer left in the plasma at these temperatures (see also Appendix~\ref{app:mesons-numeric}). As a result, a value of $X_{n}(T_0)$ is translated into $\Delta Y_{p}$ via
\begin{equation}
    \frac{\Delta Y_{p}}{Y_{p,\text{\sc sbbn}}} = \frac{\Delta X_{n}(T_{\bbn})}{X_{n,\text{\sc sbbn}}(T_{\bbn})},
    \label{eq:he-abundance-n-abundance}
\end{equation}
where $T_{\bbn}\approx 84$~keV is the temperature of the onset of nuclear reactions in SBBN~\cite{Pitrou:2018cgg}. The maximal admissible correction~\eqref{eq:he-correction-maximal} is reached for $T_{0}^{\text{min}} = 1.50$~MeV, almost independently on HNL mass (see Fig.~\ref{fig:mesons-Xn-behavior} and Appendix~\ref{app:mesons-numeric}).
Plugging $T_{0}^{\min} = 1.50$~MeV into~\eqref{eq:mesons-constraint}, we get our final limit
\begin{equation}
\tau_{N}\lesssim 0.023\s.
\label{eq:hnl-constraint-total}
\end{equation}
To obtain the bound~\eqref{eq:mesons-constraint}, we considered exclusively meson-driven $\pn$ processes for $T> T_{0}^{\text{min}}$ and only weak SBBN processes for $T<T_{0}^{\min}$. We also solved numerically the equation~\eqref{eq:Xn-equation} for the neutron abundance in the presence of both mesons-driven and SBBN $\pn$ conversion rates in Appendix~\ref{app:mesons-numeric}, and obtained constraints at the level of $0.019-0.021\s$, in perfect agreement with the bound~\eqref{eq:hnl-constraint-total}.
We have also repeated our analysis for the case of the GeV-mass scalar that mixes with the Higgs and found an excellent agreement with~\cite{Fradette:2017sdd,Fradette:2018hhl}.

Our analysis remains valid until $\tn$ reaches $\mathcal{O}(40\s)$. HNLs with longer lifetimes survive until the onset of the nuclear reactions. Mesons from their decays may then dissociate already formed nuclei, driving $Y_p$ down again, see Appendix~\ref{app:mesons-dissociation}. This effect has been analyzed in~\cite{Bondarenko:2021cpc} demonstrating that such long-lived HNLs are also excluded.

\myparagraph{Conclusion.} 
We demonstrated that HNLs with semi-leptonic decay channels significantly affect the primordial \He abundance, as  mesons from their decays drive the $\pn$ conversion rates away from their SBBN values (\textit{c.f.} \cite{Pospelov:2010cw,Fradette:2017sdd,Fradette:2018hhl}).
In order to avoid \He overproduction, mesons should disappear from the primordial plasma by $T=T_{0}^{\min}\simeq\unit[1.50]{MeV}$. 
The neutron abundance will then have enough time to relax down to its SBBN value before the onset of deuteron formation. These requirements severely constrain the parameter space of the HNLs with $0.023\s \le \tn \le 40\s$ for masses $m_N > \unit[140]{MeV}$. Accidentally, HNLs with $\tn \gtrsim 40\s$ are also excluded~\cite{Domcke:2020ety,Bondarenko:2021cpc}, but we do not consider this case here.

\begin{figure}[H]
    \begin{center}
    \includegraphics[width=0.33\textwidth]{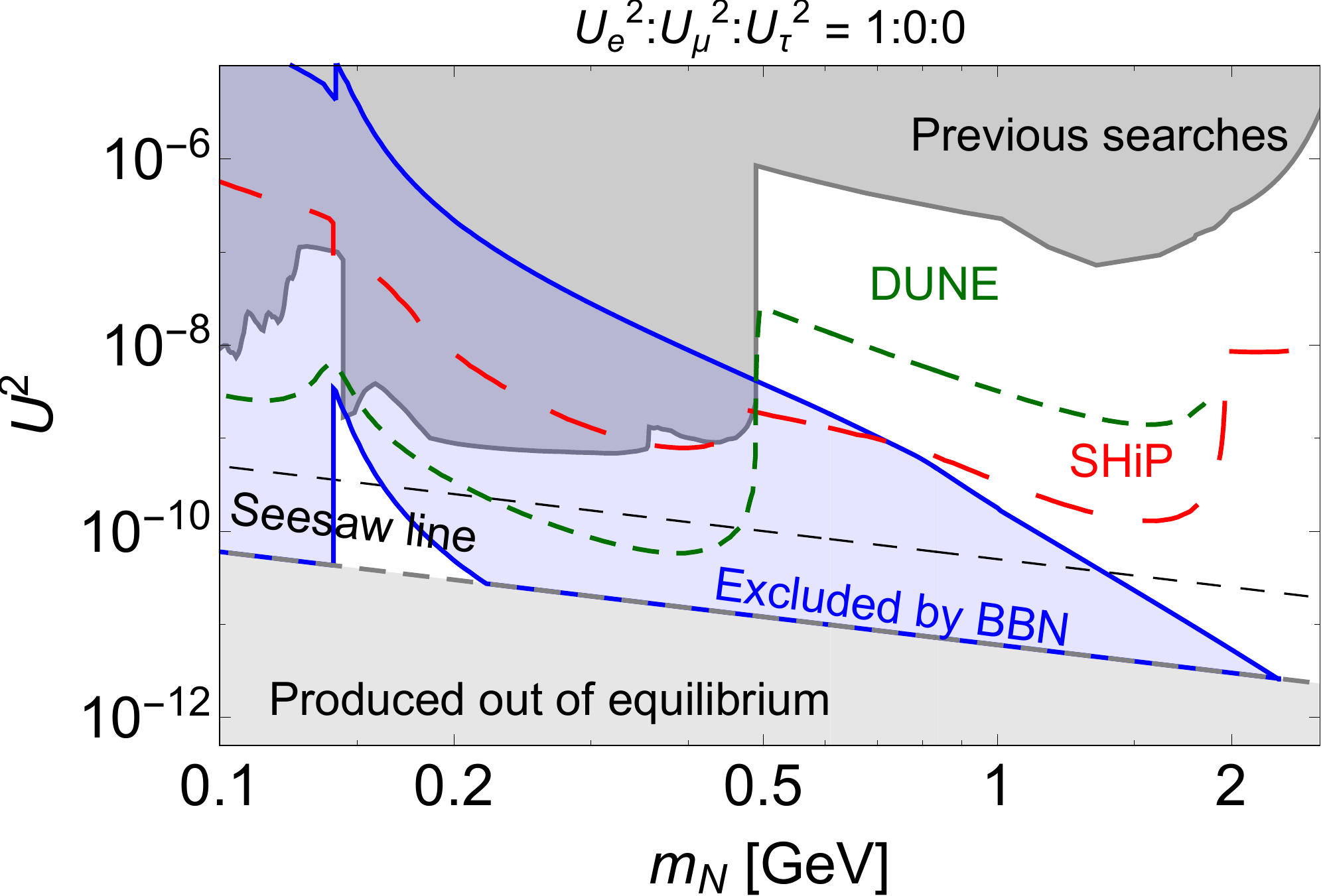}~
    \includegraphics[width=0.33\textwidth]{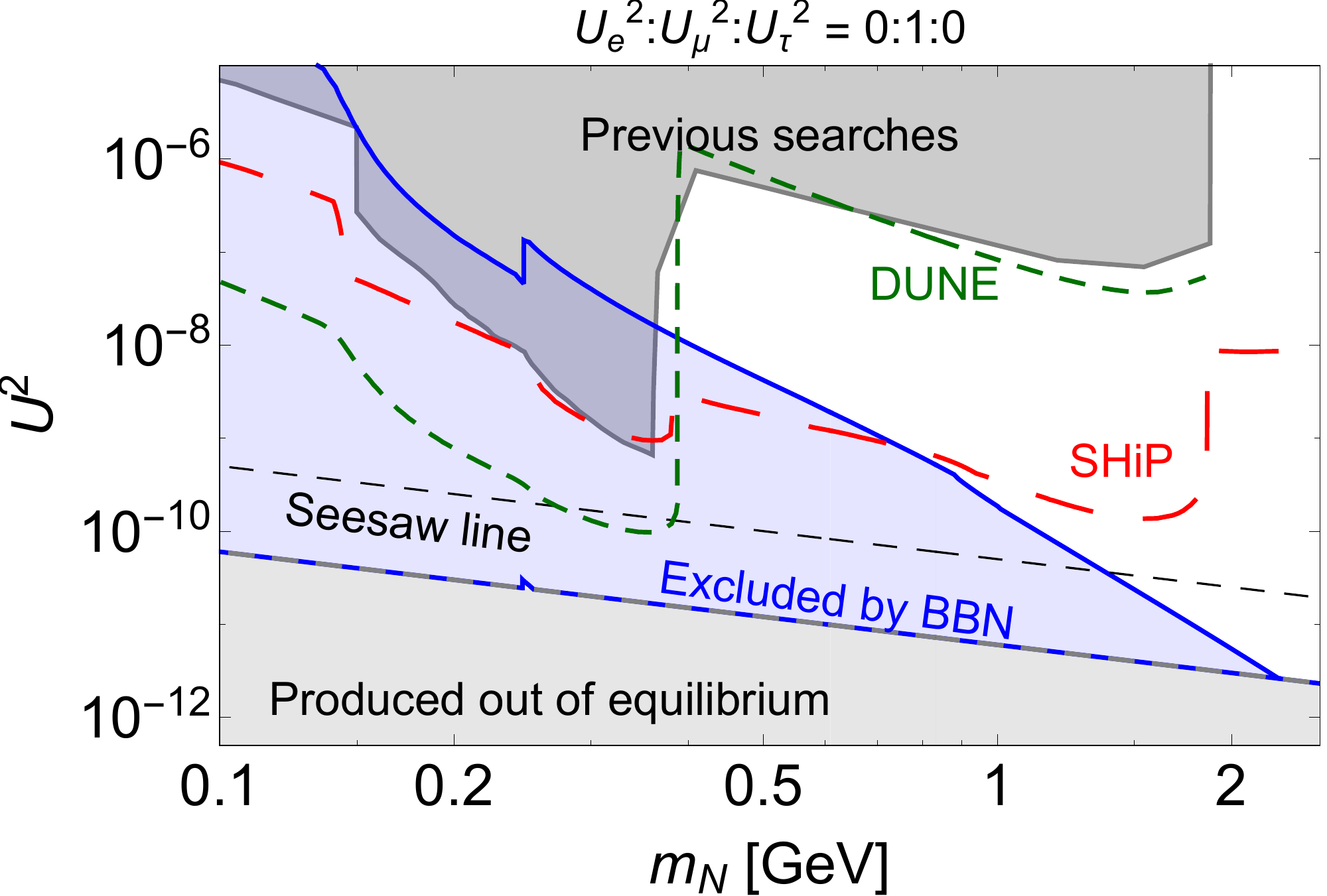}~
    \includegraphics[width=0.33\textwidth]{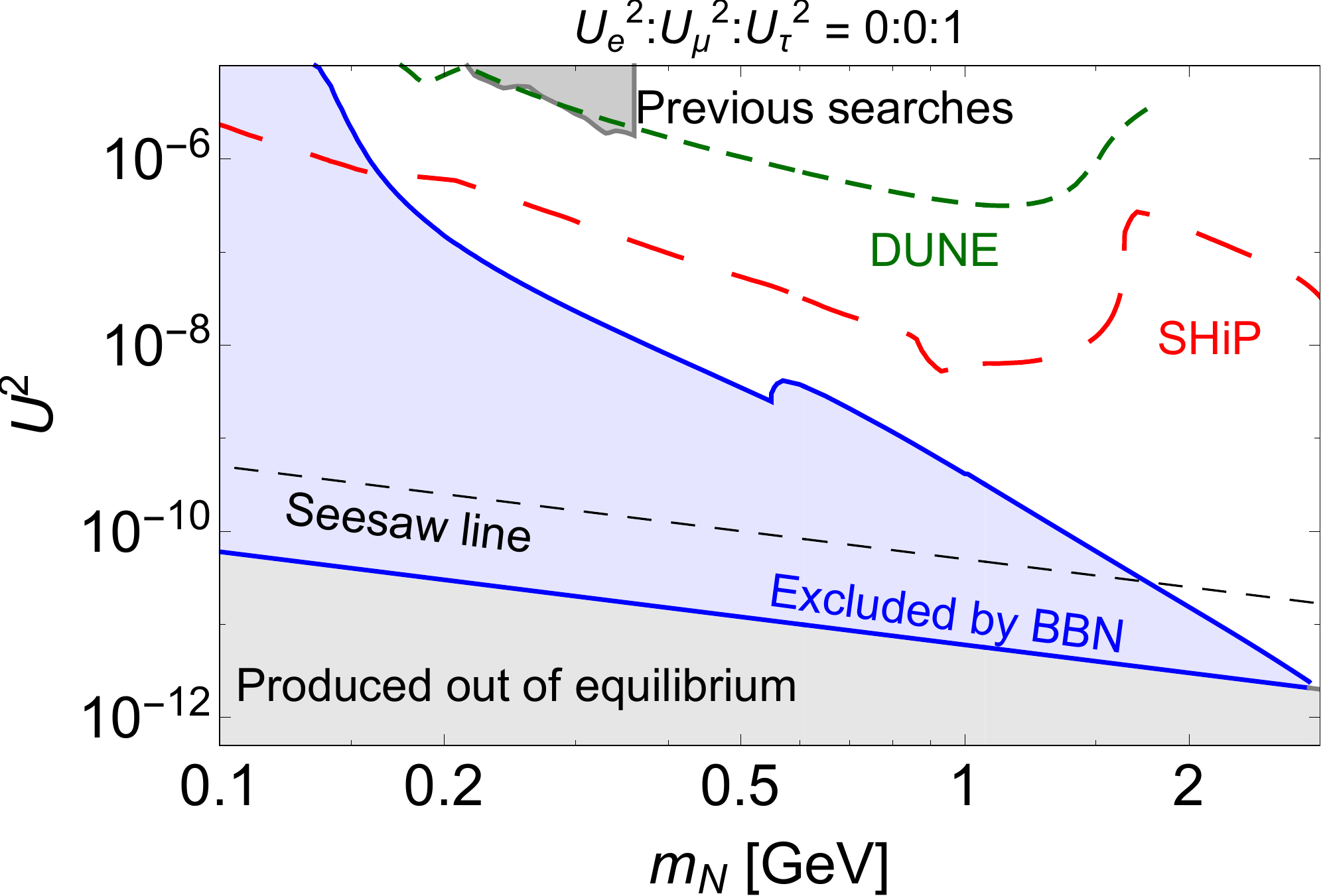}
    \end{center}
    \caption{Bounds for HNLs mixed with a particular flavor. The blue area is excluded by our present analysis combined with~\cite{bbn-large} (for HNL masses below the charged pion production threshold).
    The dark gray area denotes the excluded HNL parameter space from previous searches~\cite{Alekhin:2015byh}, including the latest NA62 search~\cite{NA62:2020mcv}. The red and greed dashed lines show the sensitivity of several future intensity frontier experiments with the highest sensitivity in the regions of interest -- SHiP~\cite{SHiP:2018xqw,Gorbunov:2020rjx} and DUNE~\cite{Ballett:2019bgd,Abi:2020evt,Coloma:2020lgy} (see~\cite{Beacham:2019nyx}). Finally, the black dashed line denotes the seesaw bound applicable if two degenerate in mass HNLs are responsible for neutrino oscillations (as in the $\nu$MSM)~\cite{Ruchayskiy:2011aa,Alekhin:2015byh}.
    The lower bound of the blue region shows the limit of the applicability of our approach: we only considered HNLs that are produced thermally and are sufficiently short-lived such that they do not change the nuclear reaction framework by their meson decay products.}
    \label{fig:hnl-constraints-intensity-frontier}
\end{figure}

We show our results for the case of two nearly-degenerate in mass HNLs that entered in thermal equilibrium and then froze out (which is reflected in both abundance calculation and decay width/patterns), as motivated by the Neutrino Minimal Standard Model (or $\nu$MSM) \cite[see e.g.][]{Boyarsky:2009ix,Eijima:2018qke,Klaric:2020lov}).
The final bounds for different mixing patterns are shown in Figs.~\ref{fig:bbn-bounds-comparison}  and~\ref{fig:hnl-constraints-intensity-frontier}.
Our constraints can be generalized to other HNL models, see e.g.~\cite{Bondarenko:2021cpc}.

Confronted with the bounds from accelerator searches, we ruled out HNLs with mass below $500$~MeV (for electron mixing) and $350$~MeV (for muon mixing). Moreover, tighter bound means that future searches at Intensity Frontier (specifically, SHiP experiment~\cite{SHiP:2018xqw}) can reach the BBN bottom line and completely rule out HNLs with the masses up to $750$~MeV, which was not the case before~\cite[see e.g.][]{Beacham:2019nyx,Strategy:2019vxc}.

The comparison  with the previous results~\cite{Ruchayskiy:2012si,Dolgov:2000jw,Sabti:2020yrt} is shown  in Fig.~\ref{fig:bbn-bounds-comparison} (right panel). Our bound~\eqref{eq:hnl-constraint-total} is a factor of $\sim 5$ stronger than the previous result~\cite{Dolgov:2000jw}. 
The recent reanalysis~\cite{Sabti:2020yrt} did not take into account the effects of mesons, therefore their results are a factor $2-3$ less conservative.

The clear qualitative effect discussed in this paper not only leads to a tighter bound on HNL lifetime and provides an reachable goal for experimental searches, but also allows for an analytic description, unusual in the realm of BBN predictions driven by sophisticated numerical codes.
\onecolumngrid 

\twocolumngrid

\myparagraph{Acknowledgments.}
We thank K.~Bondarenko, N.~Sabti for useful discussions and Yu.~Izotov for discussions related to the uncertainties of primordial Helium determination.
This project has received funding from the European Research Council (ERC) under the European Union's Horizon 2020 research and innovation programme (GA 694896) and from the Carlsberg Foundation.

\bibliographystyle{naturemag}
\providecommand{\bibinfo}[2]{\it #2}
\bibliography{bbn}

\clearpage

\onecolumngrid

\appendix 

\section{Primordial helium abundance}
\label{app:he-measurements-errors}

\begin{figure}[!h]
    \centering
    \includegraphics[width=0.5\textwidth]{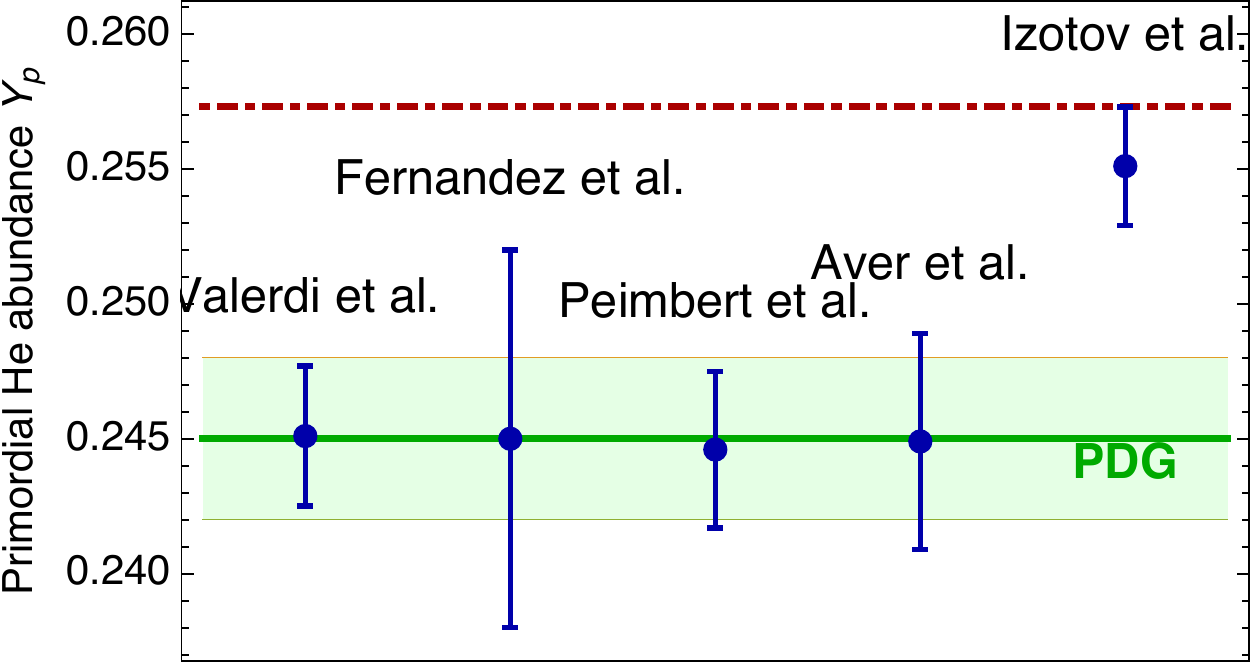}
    \caption{Measurements of $Y_{p}$ of recent works~\cite{Izotov:2014fga,Aver:2015iza, Peimbert:2016bdg,Fernandez:2018xx,Valerdi:2019beb}. The green shaded region is the PDG recommended value~\cite{Tanabashi:2018oca} (with $\pm 1 \sigma$). The gray dashed line denotes the SBBN prediction $\bar{Y}_{p} = 0.247$ from~\cite{Pitrou:2018cgg}. The red dashed-dotted line is the maximal admissible value $Y_{p,\text{max}}$ on which we base our analysis.}
    \label{fig:he-measurements-timeline}
\end{figure}

Over the last 6 years five works determined the primordial \He abundance from stellar measurements~\cite{Izotov:2014fga,Aver:2015iza, Peimbert:2016bdg,Fernandez:2018xx,Valerdi:2019beb}.
The formal statistical errors of $Y_p$ are at the level of $1-3\%$, however, the scatter between different groups is larger, see Fig.~\ref{fig:he-measurements-timeline}.

All these works determine astrophysical Helium abundance through measurements of recombination emission lines of \He and H in the metal-poor extragalactic ionized regions, then \textit{linearly} extrapolating the measurements to zero metallicity.  
Given the high precision of the results, it is important to take into account various smaller effects: including \He fluorescent emission, different ion temperatures, spatial temperature fluctuations, and others~\cite{Izotov:1999wa,Olive:2004kq}.
Additionally, while it is true that the metallicity and Helium abundance are positively correlated, the linear extrapolation to zero-metallicity may be prone to systematic uncertainties.
 
The value of $Y_{p}$ predicted within the framework of SBBN is $\bar{Y}_{p} = 0.24709\pm 0.00019$ (see, e.g.,~\cite{Pitrou:2018cgg}). The effect of mesons leads to an increase of $Y_{p}$ as compared to the SBBN value. Therefore, in order to get a conservative upper bound we assume that the maximally allowed $Y_{p}$ is given by the $1\sigma$ deviation from the maximal value predicted by~\cite{Izotov:2014fga,Aver:2015iza, Peimbert:2016bdg,Fernandez:2018xx,Valerdi:2019beb}, which is $Y_{p,\max} = 0.2573$. Note that this upper value significantly deviates from the PDG-recommended value~\cite{Tanabashi:2018oca} $Y_{p,\text{max}} = 0.248$ at $1\sigma$. This translates to the bound
\begin{equation}
    \frac{\Delta Y_{p}}{\bar{Y}_{p}} < 4.35\%
\end{equation}

\section{Evolution of HNLs in early Universe}
\label{app:hnls-population}
In this work we consider HNLs that have entered thermal equilibrium in the early Universe and then froze out. 
Below we provide necessary details of their evolution, to make our presentation more self-contained~\cite{bbn-large}. 
\begin{figure}[!h]
    \centering
    \includegraphics[width=0.5\textwidth]{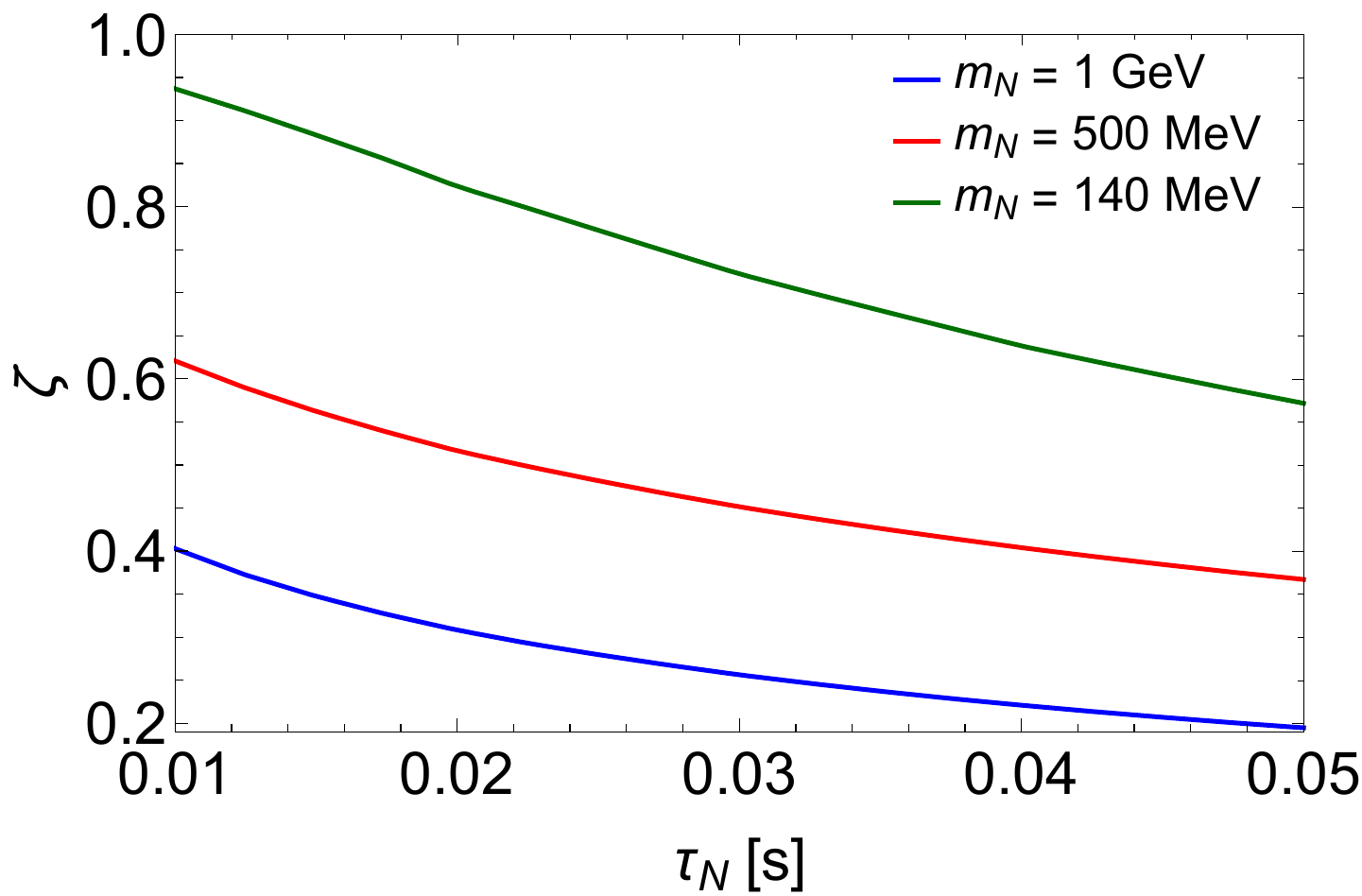}~\includegraphics[width=0.5\textwidth]{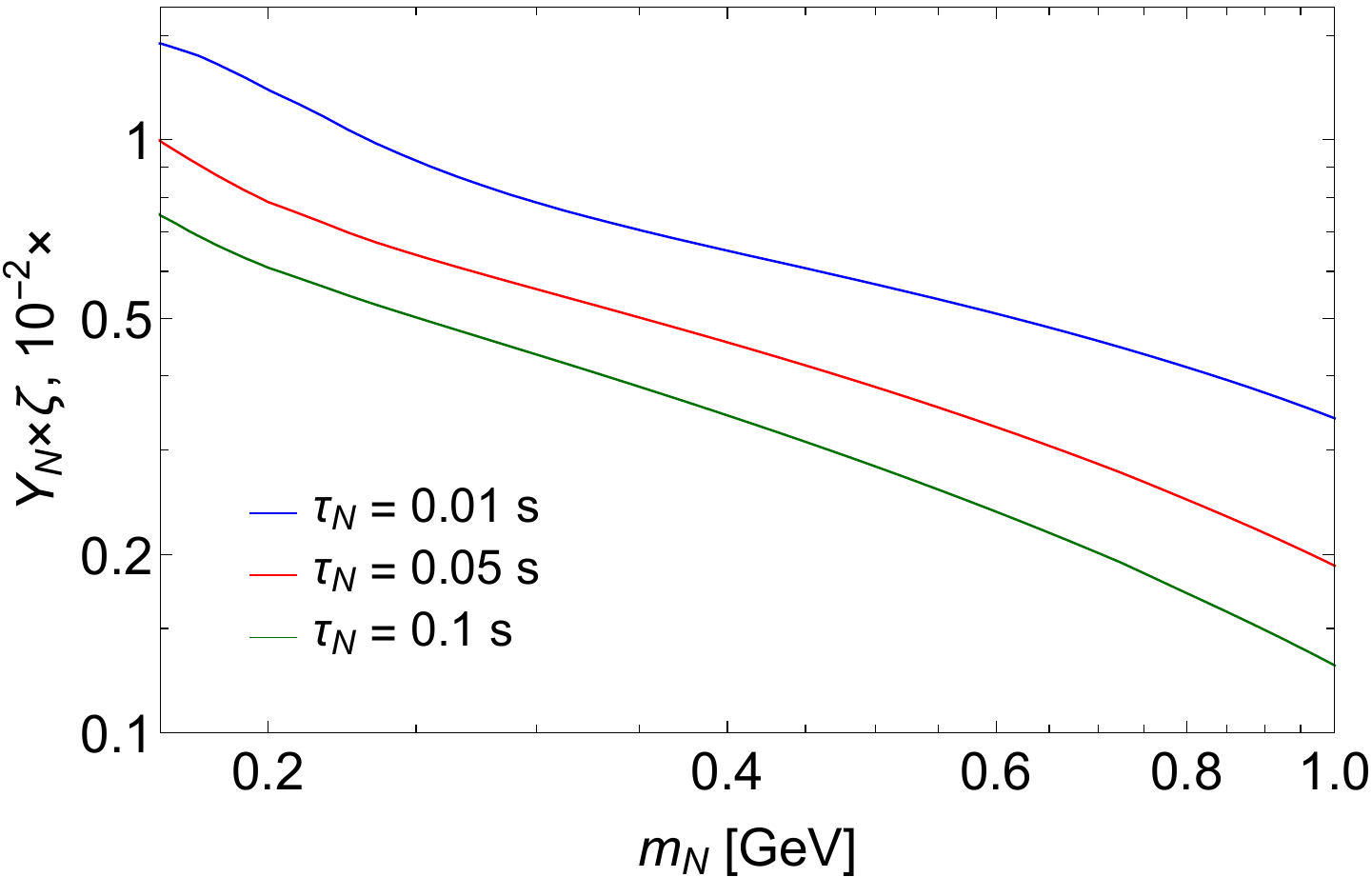}\\
    \caption{\textit{Left panel}: Dilution factor~\protect\eqref{eq:dilution-factor} for short-lived HNLs mixing with $\nu_{e}$. \textit{Right panel}: HNL abundance times dilution factor as a function of mass for particular values of the lifetime. Details of the calculation of the abundances and $\zeta$ are given in~\cite{bbn-large}. Dilution factor is calculated, when most of HNLs has decayed and do not contribute to entropy density. Note, that we define abundance at the moment of decoupling, hence it does not change with decays.}
    \label{fig:hnl-abundances-dilution}
\end{figure}
The evolution of HNLs with masses above the pion production threshold proceeds in three stages:
\begin{enumerate}
    \item The freeze-out of HNLs occurs at some temperature $T_{\dec}$ defined via
\begin{equation}
\Gamma_N^{\rm int}(T_{\dec}) \simeq H(T_{\dec}),
\label{eq:hnl-decoupling}
\end{equation}
where $\Gamma_N^{\rm int}$ is a rate of processes that keep HNLs in thermal equilibrium with the primordial plasma,\footnote{We use the matrix elements of all relevant processes with HNLs from~\cite{Sabti:2020yrt}.} and the Hubble rate $H$ corresponds to its Standard Model value (at the decoupling, HNLs with masses $m_{N}\gtrsim m_{\pi}$ and lifetimes in the range of interest contribute just a small fraction of the energy density of the Universe). Within the $\mathcal{O}(1)$ accuracy, the decoupling temperature $T_\dec$ may be estimated as
\begin{equation}
    \label{eq:Tminus}
    T_\dec \simeq T_{\nu,\rm dec}\times \begin{cases}
    \begin{aligned}
    \frac{1}{U^{2/3}} \frac1{n_{\rm int}^{1/3}}\left(\frac{g_{*}(T_{\dec})}{10.75}\right)^{1/6},&\quad&m_N \gtrsim \unit[200]{MeV},\quad \tau_N \lesssim 0.1\s\\ 
    \frac{1}{U^{2}}\frac{1}{n_{\rm int}}\left(\frac{100\text{ MeV}}{m_{N}}\right)^{2}
    \left(\frac{g_{*}(T_{\dec})}{10.75}\right)^{1/2}&\quad&m_N \lesssim \unit[200]{MeV},\quad \tau_N \lesssim 0.1\s,\\ 
    \end{aligned}
    \end{cases}
\end{equation}
where $T_{\nu,\rm dec} \approx 1.4\text{ MeV}$ is the decoupling temperature of active neutrinos, a parameter $n_{\rm int} = \frac{\Gamma_{N}^{\rm int}}{U^{2}G_{F}^{2}T^{5}}$ varies from $\simeq 1.5$ at $T\simeq \mathcal{O}(1\text{ MeV})$ to $\simeq 10$ at GeV temperatures (for Dirac HNLs). According to this estimate, for lifetimes $\tau_{N}\sim 0.01-0.1\s$, HNLs decouple while being non-relativistic for $m_N \lesssim 200$~MeV and relativistic otherwise. To improve this estimate, we solve the following equation:
\begin{equation}
    \frac{dn_{N}}{dt} + 3 H_{\text{SM}}(t)\cdot n_N = -\Gamma_N^{\rm int}(T)(n_{N}-n_{N,\text{eq}}),
    \label{eq:hnl-abundance-equation}
\end{equation}
where $n_{N,\text{eq}}$ is the number density of HNLs at equilibrium (i.e., calculated using the Fermi-Dirac distribution). Expressing the total HNL number density in terms of the abundance $Y_N$,
\begin{equation}
Y_{N} = \left( \frac{n_{N}}{s}\right)_{T = T_{\dec}},   
\label{eq:hnl-abundance}
\end{equation}
where $s = g_{*} \frac{2\pi^2}{45} T^3$ is the entropy density, we find $Y_N \simeq \frac{0.6}{g_*(T_{-})}$ in the ultra-relativistic regime and a factor of $\mathcal{O}(2)$ larger otherwise, see Fig.~\ref{fig:hnl-abundances-dilution}.

After the freeze-out, the comoving number density of HNLs changes only due to HNL decays. The physical number density thus evolves as
\begin{equation}
    n_{N}(T) =  n_{N}(T_{\dec})\cdot \left( \frac{a(T_{\dec})}{a(T)}\right)^{3}\cdot e^{-t/\tn}
    \label{eq:number-density-scaling}
\end{equation}
\item Decays of HNLs inject  energy  into the primordial plasma. This changes the time-temperature relation and the scale factor evolution as compared to SBBN. The HNL decays provide additional dilution of any decoupled relics (including  themselves) in comparison to the SBBN case:
\begin{equation}
    \zeta = \left( \frac{a_{\text{\sc sbbn}}}{a_{\text{\sc sbbn + N}}}\right)^{3} < 1,
    \label{eq:dilution-factor}
\end{equation}
where $a_{\text{\sc sbbn}}^{-1}(T) \propto g_{*}^{1/3}T$ is the scale factor in SBBN, and the scale factors are evaluated at  times $t\gg \tn$. To calculate $\zeta$, we solve the Friedmann equation under an assumption that neutrinos are in perfect equilibrium and neglecting the mass of electrons:
\begin{equation}
\begin{aligned}
    H^2(t) = &\frac{1}{M^{2}_{\text{Pl}}}\frac{8\pi}{3}\biggl[\rho_{\text{rad}} + m_N\cdot  n_N(T)\biggr], \\ 4\frac{\rho_{\text{rad}}}{T}\frac{dT}{dt} = &\frac{m_N n_N(t)}{\tau_N}  -4 H(t)\cdot\rho_{\text{rad}},
\end{aligned}
\label{eq:friedmann}
\end{equation}
where the number density of HNLs is given by Eq.~\eqref{eq:number-density-scaling}. This is a reasonable assumption, since most of the HNLs with lifetimes $\tn\ll 0.1\s$ decay much earlier than neutrinos decouple.
 
\item Strong meson effects mean that we need to trace the number of HNLs even at times $t\gg \tn$:
\begin{equation}
    n_{N}(t\gg \tn) = n_{N}(T_{-})\cdot \left(\frac{a_{\text{\sc sbbn}}(T_{-})}{a_{\text{\sc sbbn}}(T)}\right)^{3}\cdot \zeta \cdot e^{-t(T)/\tn} \approx 0.4 Y_{N}\cdot g_{*,\text{\sc sbbn}}\ T^{3}\cdot \zeta \cdot e^{-t(T)/\tn},
    \label{eq:number-density-exponential-tail}
\end{equation}
where $t(T)$ is the same as in SBBN.\footnote{At times $t\gg \tn$, the time-temperature relation differs from SBBN only by the value of $N_{\eff}$. However, the latter may change only if neutrinos are not in perfect equilibrium, and hence $t(T)$ is the same as in SBBN for lifetimes $\tn \ll 0.1$~s.} Because of the suppression, the effect of this population on the expansion of the Universe may be neglected. However, this exponential tail still may produce mesons in amounts sufficient to change the dynamics of the $n/p$ ratio.
\end{enumerate}
The values of the HNL abundance and the dilution factor versus its mass and lifetime are given in Fig.~\ref{fig:hnl-abundances-dilution}.

\section{Hadronic decays of HNLs}
\label{app:hadronic-hnl-decays}
\begin{figure}[!t]
    \centering
    \includegraphics[width=0.7\textwidth]{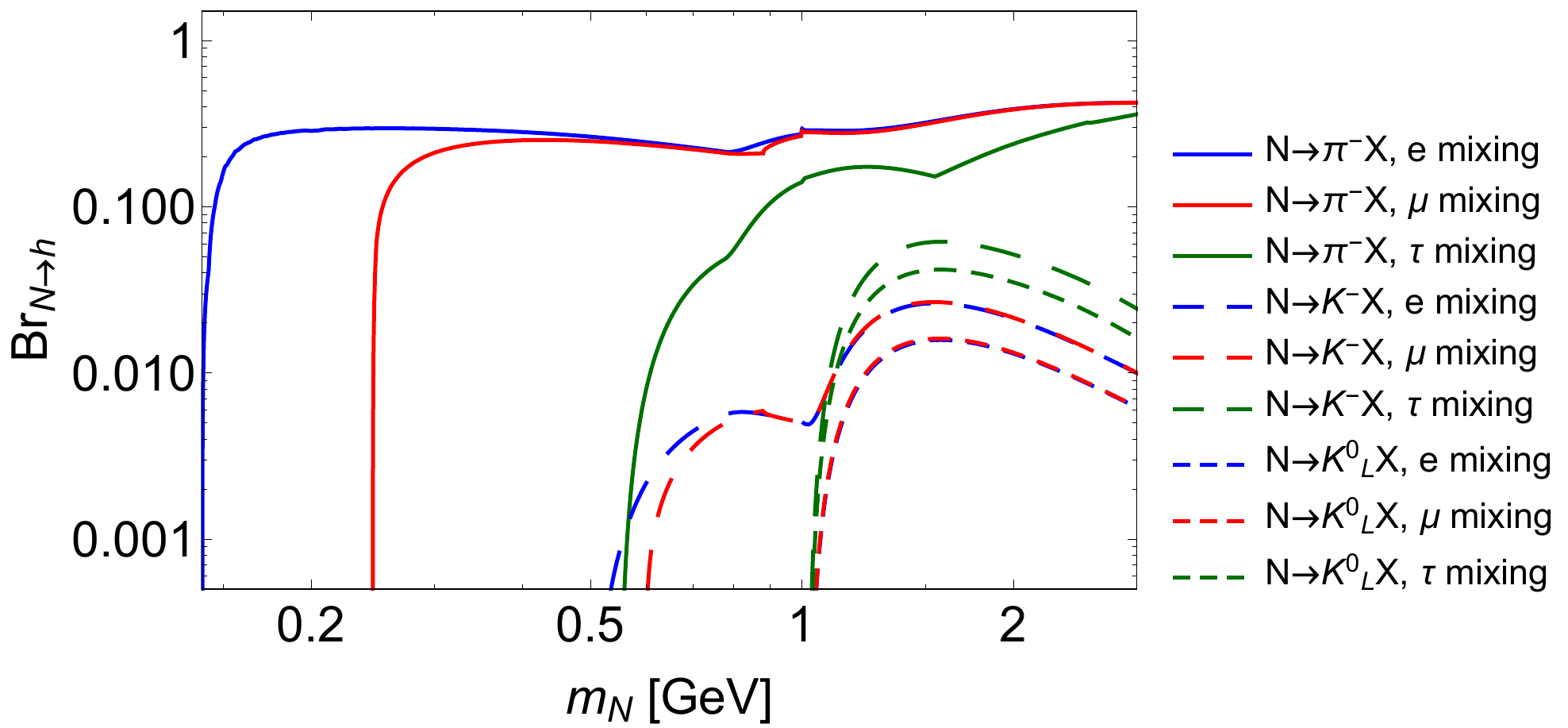}
    \caption{Branching ratios of HNL decays  into mesons $h = \pi^{-}, K^{-}, K^{0}_{L}$. Secondary decays are also included (see text for details).}
    \label{fig:HNLs-decay-br-mesons}
\end{figure}

In this work, we consider a pair of HNLs, degenerate in mass and having similar mixing angles. Two such HNLs form a single quasi-Dirac fermion \cite{Shaposhnikov:2006nn,Kersten:2007vk}. The abundance of a meson $h$ produced from such HNLs is proportional to the quantity $Y_{N}\cdot \Br_{N\to h}$. The mass dependence of $\Br_{N\to h}$ for different mesons $h$ and mixing patterns is shown in Fig.~\ref{fig:HNLs-decay-br-mesons}. 
We are interested only in the abundances of light mesons (pions and kaons) and for HNL masses well above pion/kaon thresholds we should account for ``secondary mesons''.
This is discussed below, mainly following~\cite{Bondarenko:2018ptm}.
\myparagraph{Decays into pions.} 
In the case of the pure $e/\mu$ mixings, the charged pion production threshold corresponds to $\mn = m_{\pi} + m_{l}$, where $l = e/\mu$. For $\tau$ mixing, the similar charged current-mediated channel opens up only at $m_{N}=m_{\tau}+m_{\pi}\simeq 1.9\text{ GeV}$. However, for all types of mixings charged pions may appear as secondary particles in decays of neutral mesons,
\begin{equation}
N \to h^{0} + \nu_\alpha, \quad h^{0}\to \pi^{\pm}+X,\quad \text{where} \quad h^{0} = \rho^{0},\eta^{0},\eta',\omega^{0},\phi
\end{equation}
Therefore, for $\tau$ mixing charged pions may appear at masses $\mn \ge m_{\eta^0}$. We use the branching ratios $\Br_{\eta^{0}\to \pi^{\pm}X} \approx 0.27$, $\Br_{\rho^{0,\pm}\to \pi^{\pm}X} \approx 1$~\cite{Tanabashi:2018oca}.

Above $\mn \simeq 1$~GeV, decays of HNLs into pions cannot be approximated by single meson decays. Indeed, decays of GeV mass range HNLs are similar to decays of $\tau$ lepton~\cite{Bondarenko:2018ptm}, whereas for the latter hadronic decays are dominated by multi-pion channels~\cite{Tanabashi:2018oca}. We estimate the width of multi-pion decays as the difference between the total width into quarks and the width into single mesons:
\begin{equation}
 \Gamma_{N\to n\pi} = \Gamma_{N\to\text{quarks}}-\sum_{h = \pi,K,\rho,\dots}\Gamma_{N\to hX}
\end{equation}
For multiplicities $\mathcal{N}$ of decays of HNLs into charged pions $N\to \pi^{\pm}$ (i.e., the amount of $\pi^{\pm}$ per multi-hadronic decay of HNLs), we will use multiplicities for multihadronic decays of $\tau$ leptons. Namely, $\mathcal{N}_{N\to\pi^+} = \mathcal{N}_{\tau^{+}\to \pi^{+}}\approx 1.35$, $\mathcal{N}_{N\to \pi^-} = \mathcal{N}_{\tau^{+}\to \pi^{-}}\approx 0.34$. The effective branching into $\pi^{-}$ from multi-pion decays is
\begin{equation}
    \Br_{N\to \pi^{-}}^{\text{multi-pion}} = \mathcal{N}_{N\to\pi^{-}}\cdot \frac{\Gamma_{N\to n\pi}}{\Gamma_{N}}, \qquad \Br_{\bar{N}\to \pi^{-}}^{\text{multi-pion}} = \mathcal{N}_{\bar{N}\to \pi^{-}}\cdot \frac{\Gamma_{N\to n\pi}}{\Gamma_{N}} 
\end{equation}
Since the bound on the meson driven \pn conversion is only logarithmically sensitive to the value of $\Br_{N\to \pi^{\pm}}$, our results depend on these assumptions weakly. 

\myparagraph{Decays into kaons.} Below $m_{N}= m_{\phi}$, charged kaons may appear only through the mixing with $e/\mu$ in the process $N\to K^{-}l$. This decay is Cabibbo suppressed~\cite{Bondarenko:2018ptm} and almost two orders of magnitude smaller than into pions. Neutral kaons appear only in the final states with three or more particles (such as $N \to K^0 + \bar K^0 + \nu_\alpha$ and $N\to K^{+} + \bar K^0 + \ell^-$, etc).

HNLs heavier than $\phi$ meson may produce both charged and neutral kaons via decays $N\to \phi \nu, \phi \to KK$. We assume that $K^{0}$ contains equal admixtures of $K^{0}_{L}$ and $K^{0}_{S}$, i.e.\ $\Br_{N \to K^{0}_{L}} = \Br_{N\to K^{0}}/2$. We use the branching ratios $\Br_{\phi\to K^{-}}\approx 0.5$, $\Br_{\phi\to K^{0}_{L}}\approx 0.34$~\cite{Tanabashi:2018oca}.

\section{Changes in $\pn$ rates due to the presence of mesons}
\label{app:mesons}
In this section, we provide details on our estimate of the effect of mesons on BBN.
\subsection{Processes and cross-sections}
\label{app:mesons-xsection}
\paragraph{Pions.}
The threshold-less processes with charged pions are
\begin{equation}
    \pi^{-}+p \rightarrow n + \pi^{0}/\gamma, \quad \pi^{+}+n \rightarrow p+ \pi^{0}.
    \label{eq:pions-processes}
\end{equation}
The cross-sections at threshold are~\cite{Kohri:2001jx}
\begin{equation}
\langle\sigma_{p\to n}^{\pi^{-}}v\rangle \approx 4.3\cdot 10^{-23}F_{c}^{\pi}(T)\text{ m}^{3}/s, \quad \frac{\langle\sigma_{p\to n}^{\pi^{-}}v\rangle}{\langle\sigma_{n\to p}^{\pi^{+}}v\rangle}\approx 0.9\ F_{c}^{\pi}(T),
\label{eq:pions-cross-sections}
\end{equation}
where $F_{c}^{h}$ is the Sommerfeld enhancement of the cross-section due to presence of two oppositely charged particles in the in-state:
\begin{equation}
    F^{h}_{c} = \frac{x}{1+e^{-x}},\quad \text{where} \quad x = \frac{2\pi \alpha_{\EM}}{v_{e}},
    \label{eq:sommerfeld-enhancement}
\end{equation}
where $v_{e} \approx \sqrt{\frac{T}{m_{h}}}+\sqrt{\frac{T}{m_{p}}}$ is the relative velocity between a nucleon and a meson. $F_{c}$ is of order of one at $T\simeq 1\text{ MeV}$.

\paragraph{Kaons.}
The threshold-less $n\leftrightarrow p$ conversions driven by kaons are
\begin{equation}
\begin{aligned}
K^{-}+p &\to \Sigma^{\pm/0}/\Lambda+\pi^{\mp/0}/\pi^{0} \to n + 2\pi, \\ K^{-}+n &\to \Sigma^{-/0}/\Lambda+\pi^{0/-}/\pi^{-} \to n + 2\pi, \\ \bar{K}^{0}_{L} + p &\to \Sigma^{0/+}/\Lambda +\pi^{+/0}/\pi^{+}\to n + 2\pi, \\ \bar{K}^{0}_{L} + n &\to \Sigma^{\pm/0}/\Lambda +\pi^{\mp/0}/\pi^{0}\to p + 2\pi,
\end{aligned}
\label{eq:kaons-processes}
\end{equation}
where $\Lambda,\Sigma$ are the lightest strange hadronic resonances~\cite{Pospelov:2010cw}. 

Their effect is similar to the one of pions, but with small differences: \textit{(i)} cross-sections of above reactions are higher than the cross-sections of~\eqref{eq:mesons-processes}\footnote{The reason is that these reactions have higher available phase space and go through hadronic resonances.}, \textit{(ii)} there is no isotopic symmetry - $K^{+}$ mesons do not contribute to \pn conversion, since there are no threshold-less processes $n + K^{+} \to p + X$. Indeed, the process $n+K^{+} \to p + K^{0}$ has the threshold $Q \approx 2.8$~MeV, while the threshold-less processes going through s-quark resonances, similar to~\eqref{eq:kaons-processes}, would require resonances with negative strangeness and positive baryon number, that do not exist, \textit{(iii)} neutral kaons do not lose the energy before decaying (however, we follow~\cite{Pospelov:2010cw} and approximate the cross-sections by threshold values). 

The threshold cross-sections are
\begin{equation}
\langle\sigma_{p\to n}^{K^{-}}v\rangle \approx 9.6\cdot 10^{-22}F_{c}^{K}(T)\text{ m}^{3}/s, \quad \frac{\langle\sigma_{p\to n}^{K^{-}}v\rangle}{\langle\sigma_{n\to p}^{K^{-}}v\rangle}\approx 2.46\ F_{c}^{K}(T),
\label{eq:charged-kaons-cross-sections}
\end{equation}
\begin{equation}
    \langle \sigma_{p\to n}^{K^{0}} v\rangle \approx 1.95\cdot 10^{-22}\text{ m}^{3}/\text{s}, \quad \frac{\langle\sigma_{p\to n}^{K^{0}_{L}}v\rangle}{\langle\sigma_{n\to p}^{K^{0}_{L}}v\rangle} \approx 0.41.
\label{eq:neutral-kaons-cross-sections}
\end{equation}

\paragraph{Conversion probabilities.} A probability for a meson $h$ to convert $p\leftrightarrow n$ before decaying is given by
\begin{equation}
    P_{\text{conv}}^{h} \approx \frac{\langle\sigma_{p\leftrightarrow n}^{h}v\rangle n_{B}}{\Gamma^{h}_{\decay}},
\end{equation}
where $\Gamma^{h}_{\decay}$ is the decay width and $n_{B}$ is the baryon number density. The decay widths of mesons are~\cite{Tanabashi:2018oca}
\begin{equation}
\Gamma^{\pi^{\pm}}_{\text{decay}}\approx 3.8\cdot 10^{7}\s^{-1}, \quad \Gamma^{K^{-}}_{\text{decay}}\approx 8.3\cdot 10^{7}\s^{-1}, \quad \Gamma^{K^{0}_{L}}_{\text{decay}}\approx 2\cdot 10^{7}\s^{-1}
\label{eq:mesons-decay-widths}
\end{equation}

Using~\eqref{eq:pions-cross-sections},~\eqref{eq:charged-kaons-cross-sections},~\eqref{eq:mesons-decay-widths}, for the $p\to n$ conversion probabilities we obtain
\begin{equation}
    P_{\text{conv}}^{\pi^{-}}(T) \approx 2.5\cdot 10^{-2}\left( \frac{T}{1\text{ MeV}}\right)^{3}, \quad P_{\text{conv}}^{K^{-}}(T) \approx 2.8\cdot 10^{-1}\left( \frac{T}{1\text{ MeV}}\right)^{3}, \quad P_{\text{conv}}^{K^{0}_{L}}(T) \approx 1.6\cdot 10^{-1}\left( \frac{T}{1\text{ MeV}}\right)^{3}
    \label{eq:conversion-probabilities}
\end{equation}
The largeness of the probabilities is caused by the fact that the decay of mesons proceeds through weak interactions, while the $p\leftrightarrow n$ conversion is mediated by strong interactions. In particular, at $T\gtrsim 2\text{ MeV}$ kaons participate in the conversion faster than they decay.

\subsection{Numeric study}
\label{app:mesons-numeric}
To verify the analytic estimate~\eqref{eq:hnl-constraint-total}, we numerically solve equation for the neutron abundance $X_n$, where we include both weak conversion \pn processes and the meson driven processes~\eqref{eq:pions-processes}-\eqref{eq:kaons-processes}. The system of equations has the form
\begin{equation}
\begin{cases}
 \frac{X_{n}}{dt} = \left(\frac{dX_{n}}{dt}\right)_{\text{SM}}+\left(\frac{dX_{n}}{dt}\right)_{\pi}+\left(\frac{dX_{n}}{dt}\right)_{K^{-}}+\left(\frac{dX_{n}}{dt}\right)_{K^{0}_{L}},
\\ \frac{dn_{\pi^{-}}}{dt} = n_{N}\frac{\Br_{N\to \pi^{-}}}{\tau_{N}} - \Gamma^{\pi^{-}}_{\text{decay}}n_{\pi^{-}} - \langle \sigma_{p\to n}^{\pi^{-}} v\rangle (1-X_{n})n_{B} n_{\pi^{-}},\\ \frac{dn_{\pi^{+}}}{dt} = n_{N}\frac{\Br_{N\to \pi^{+}}}{\tau_{N}} - \Gamma^{\pi^{+}}_{\text{decay}}n_{\pi^{+}} - \langle \sigma_{n\to p}^{\pi^{+}} v\rangle X_{n}n_{B}n_{\pi^{+}}, \\ \frac{dn_{K^{-}}}{dt} = n_{N}\frac{\Br_{N\to K^{-}}}{\tau_{N}} - \Gamma^{K^{-}}_{\text{decay}}n_{K^{-}} - \langle \sigma_{p\to n}^{K^{-}} v\rangle (1-X_{n})n_{B} n_{K^{-}} - \langle \sigma_{n\to p}^{K^{-}} v\rangle X_{n}n_{B} n_{K^{-}}, \\ \frac{dn_{K^{0}_{L}}}{dt} = n_{N}\frac{\Br_{N\to K^{0}_{L}}}{\tau_{N}} - \Gamma^{K^{0}_{L}}_{\text{decay}}n_{K^{0}_{L}} - \langle \sigma_{p\to n}^{K^{0}_{L}} v\rangle (1-X_{n})n_{B} n_{K^{0}_{L}} - \langle \sigma_{n\to p}^{K^{0}_{L}} v\rangle X_{n}n_{B} n_{K^{0}_{L}}
\end{cases}                
\label{eq:meson-population-evolution}
\end{equation}
Here the quantities
\begin{equation}
\label{eq:Xn_meson_rates}
\begin{aligned}
\left(\frac{dX_{n}}{dt}\right)_{\pi} = (1-X_{n})n_{\pi^{-}}\langle \sigma_{p\to n}^{\pi^{-}}v\rangle-X_{n}n_{\pi^{+}}\langle \sigma_{n\to p}^{\pi^{+}}v\rangle, \\
\text{and}\\
\left(\frac{dX_{n}}{dt}\right)_{K} = (1-X_{n})n_{K}\langle \sigma_{p\to n}^{K}v\rangle-X_{n}n_{K}\langle \sigma_{n\to p}^{K}v\rangle
\end{aligned}
\end{equation}
are the rates of change of $X_{n}$ due to different mesons ($K = K^{-}/K^{0}_{L}$);  $n_{B}$ is the baryon number density $n_{B} = \eta_{B}n_{\gamma}$. In equations for the number density of mesons $n_{h}$, the first term comes from HNLs, the second due to decays of mesons and the last term is due to $p\leftrightarrow n$ conversion. The time-temperature relation and the scale factor dynamics are provided by the solution of Eq.~\eqref{eq:friedmann}, and the HNL number density may be obtained using Eq.~\eqref{eq:number-density-scaling}.

During times $t_{\text{eq}}\simeq (\Gamma^{h}_{\text{decay}})^{-1}\sim 10^{-8}\s$, which are small in comparison to any other time scale in the system, the solution for $n_{h}$ reaches the dynamical equilibrium:
\begin{equation}
    n_{\pi^{-}} = \frac{n_{N}\cdot \Br_{N\to \pi^{-}}}{\tau_{N}(\Gamma^{\pi^{-}}_{\text{decay}}+\langle \sigma_{p\to n}^{\pi^{-}} v\rangle (1-X_{n})n_{B})}, \quad n_{\pi^{+}} = \frac{n_{N}\cdot \Br_{N\to \pi^{+}}}{\tau_{N}(\Gamma^{\pi^{+}}_{\text{decay}}+\langle \sigma_{n\to p}^{\pi^{+}} v\rangle (1-X_{n})n_{B})},
\label{eq:meson-instant-density}
\end{equation}
\begin{equation}
    n_{K} = \frac{n_{N}\cdot \Br_{N \to K}}{\tau_{N}(\Gamma^{K}_{\text{decay}}+\langle \sigma_{p\to n}^{K} v\rangle (1-X_{n})n_{B}+ \langle \sigma_{n\to p}^{K} v\rangle X_{n}n_{B})},
    \label{eq:kaon-instant-density}
\end{equation}
where $K = K^{-}/K^{0}_{L}$.

\begin{figure}[!h]
    \centering
    \includegraphics[width=0.33\textwidth]{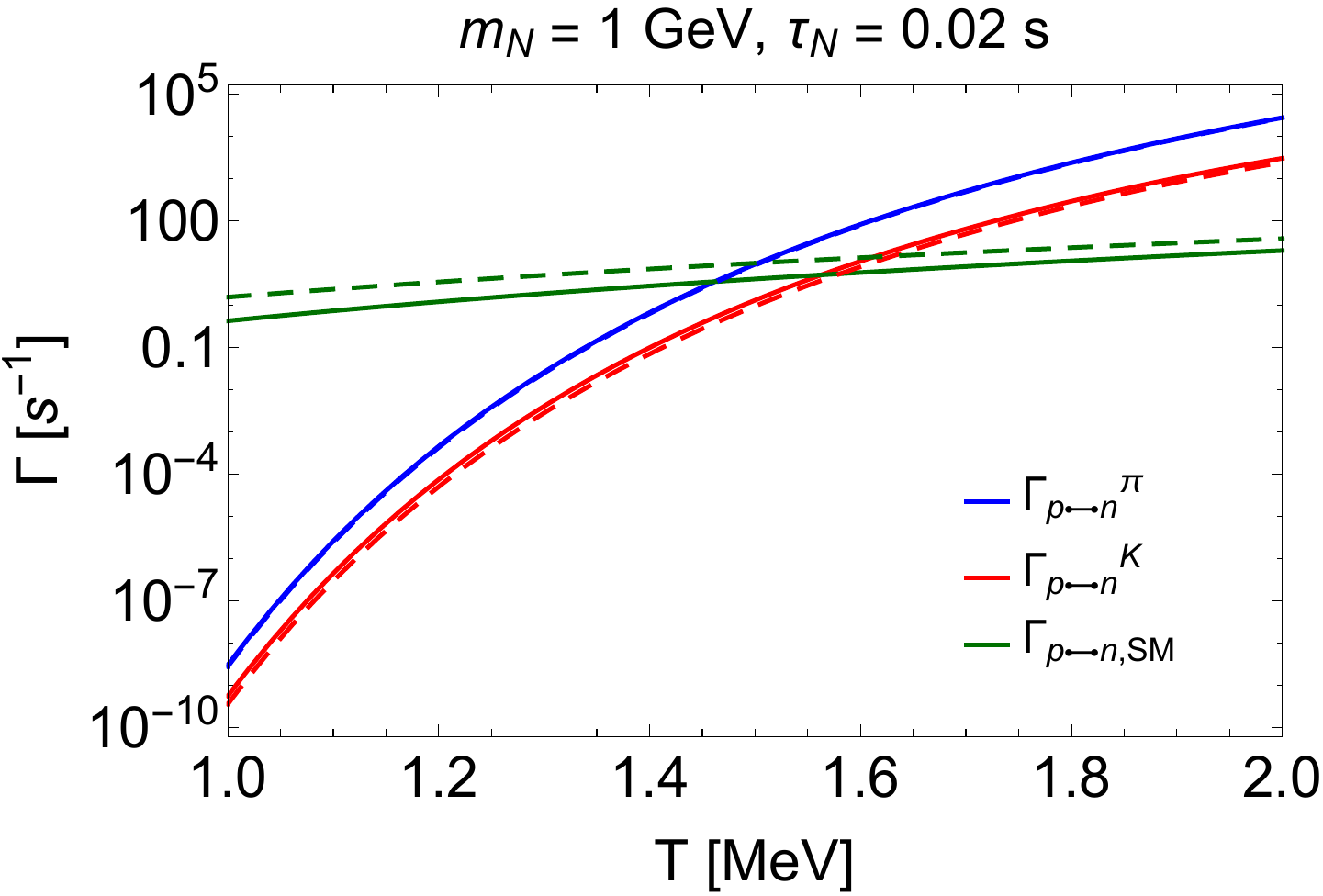}~\includegraphics[width=0.33\textwidth]{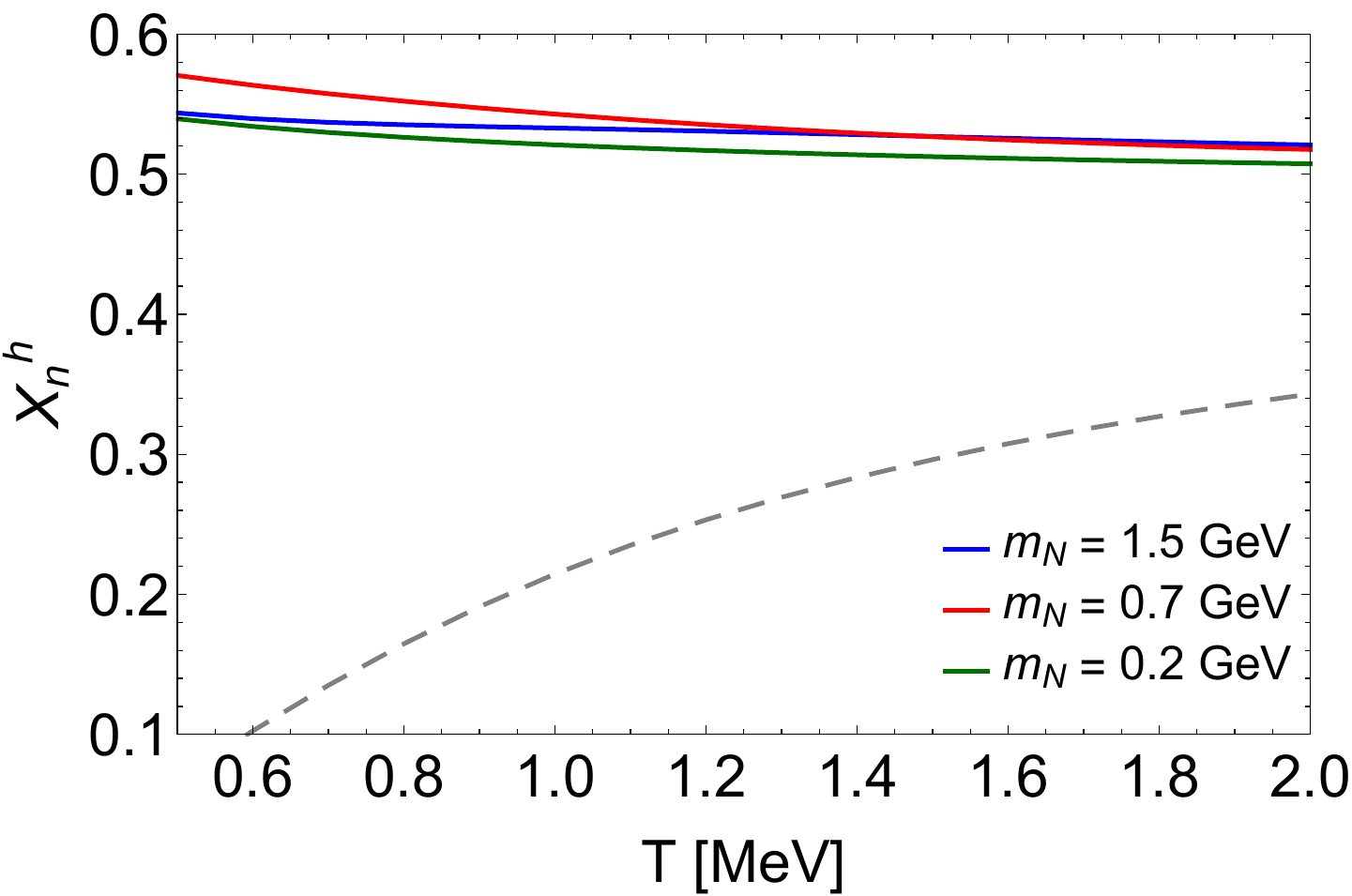}~\includegraphics[width=0.33\textwidth]{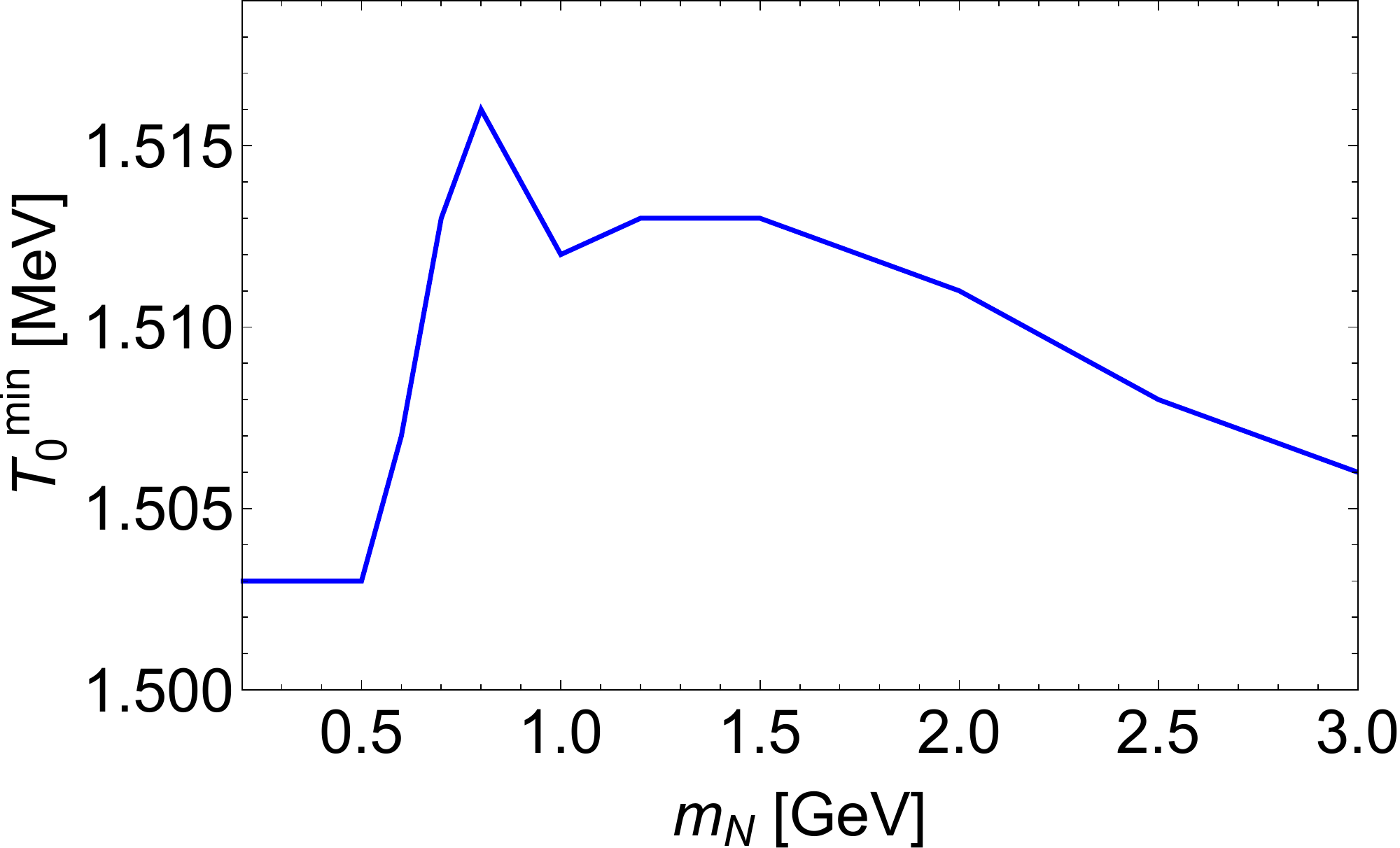}
    \caption{\textit{Left panel}: the behavior of the $p\to n$ (solid lines) and $n\to p$ (dashed lines) conversion rates in the case of pion and kaon driven conversions and SBBN. We consider HNLs mixing with $e$ flavor, mass $\mn = 1\gev$ and lifetime $\tn = 0.02\s$ as an example. \textit{Middle panel}: the temperature dependence of the neutron abundance  $X_{n}$ assuming that its evolution is completely dominated by the meson driven \pn conversions. We consider HNLs mixing with $e$ flavor and different masses: $m_{N} =200$~MeV (only pions are present), $\mn = 700$~MeV (pions and charged kaons are present), $\mn = 1.5$~GeV (pions, charged and neutral kaons are present). The dashed gray line denotes the value of the neutron abundance  at equilibrium in SBBN. \textit{Right panel}: the HNL mass dependence of the temperature $T_{0}^{\min}$.}
    \label{fig:mesons-conversion-rates-comparison}
\end{figure}

Therefore, we solve a single equation 
\begin{equation}
\frac{X_{n}}{dt} = \left(\frac{dX_{n}}{dt}\right)_{\text{SM}}+\left(\frac{dX_{n}}{dt}\right)_{\pi}+\left(\frac{dX_{n}}{dt}\right)_{K^{-}}+\left(\frac{dX_{n}}{dt}\right)_{K^{0}_{L}}.
\end{equation}
where we use meson number densities given by Eqs.~\eqref{eq:meson-instant-density} and \eqref{eq:kaon-instant-density} in the meson-driven conversion rates~\eqref{eq:Xn_meson_rates}. The results are shown in Fig.~\ref{fig:mesons-conversion-rates-comparison}.  Our main result is the right panel of Fig.~\ref{fig:mesons-conversion-rates-comparison} -- it shows that the value $T_0^{\text{min}}\simeq \unit[1.50]{MeV}$ and that its variation as a function of the HNL mass is within $\pm 1\%$.

With the help of Eqs.~\eqref{eq:pions-cross-sections},~\eqref{eq:charged-kaons-cross-sections}, we obtain the value of the neutron abundance driven solely by a given meson $h$. As long as $T\gtrsim T_{0}$ (see Eq.~\eqref{eq:meson-constraint-parametric} and left panel of Fig.~\ref{fig:mesons-conversion-rates-comparison}), the weak interaction processes may be completely neglected, and the resulting $X_n$ are given by
\begin{equation}
X_{n}^{\pi^{\pm}}= \frac{\langle\sigma_{p\to n}^{\pi^{-}} v\rangle \cdot n_{\pi^{-}}}{\langle\sigma_{p\to n}^{\pi^{-}} v\rangle \cdot n_{\pi^{-}} + \langle\sigma_{n\to p}^{\pi^{+}} v\rangle \cdot n_{\pi^{+}}} \approx \frac{0.9F_{c}^{\pi}(T)}{1+0.9F_{c}^{\pi}(T)}, \quad X_{n}^{K^{-}}\approx \frac{2.46F_{c}^{K}}{2.46F_{c}^{K}+1}, \quad X_{n}^{K^{0}_{L}}\approx 0.32
\label{eq:Xneq-mesons}
\end{equation}
The values of $X_{n}^{\pi^{-}/K^{-}}$ grow with the decrease of the temperature due to the growth of the Coulomb factor $F_{c}$, which enhances the rate of the $p\to n$ process.

The quantities~\eqref{eq:Xneq-mesons} provide us the qualitative estimate of the value of $X_{n}$ in presence of different mesons, Fig.~\ref{fig:mesons-conversion-rates-comparison}. Below the kaon production threshold, $X_{n}^{h} = X_{n}^{\pi^{\pm}}$. At larger masses, in order to find $X_{n}^{h}$ we need to set the whole right hand-side of Eq.~\eqref{eq:meson-population-evolution} to zero. Below the $K^{0}_{L}$ production threshold (which occurs at $m_{N}=m_{\phi}$), the value of $X_{n}^{h}$ grows, since charged kaons tend $X_{n}$ to higher values than $X_{n}^{\pi^{-}}$. Above the neutral kaon production threshold, the ratio $\Br_{N\to K^{-}}/\Br_{N\to \pi^{-}}$ increases (Fig.~\ref{fig:HNLs-decay-br-mesons}) and $X_{n}^{h}$ grows further. However, kaons $K^{0}_{L}$, that are present in small amounts, somewhat diminish this growth.

The value of $X_{n}^{h}(m_{N})$ provide us the mass dependence of $T_{0}^{\min}(m_{N})$, which is the smallest temperature allowed by observations (c.f.\ Fig.~\ref{fig:mesons-Xn-behavior}). We show it in Fig.~\ref{fig:mesons-conversion-rates-comparison} (right panel).

\subsection[SBBN evolution at low $T$]{SBBN evolution at $T < T_0^{\text{min}}$}
\label{app:lowT}

If HNLs disappear from the plasma before neutrinos froze out, the evolution of the neutron abundance and subsequent nuclear reactions proceed exactly as in SBBN case (albeit with modified initial value of $X_n$ at $T = T_0^{\text{min}}$).

Indeed, the onset of nuclear reactions is determined by the dynamical balance between reactions of deuterium synthesis and dissociation. This balance depends on the value of $\eta_{B}$. The latter gets diluted by the factor $\zeta$ due to decays of HNLs, see Section~\ref{app:hnls-population}. However, we fix $\eta_{B}$ at the beginning of nuclear reactions to be the same as measured by CMB. This of course means that $\eta_B$ has been $\zeta^{-1}$ times higher before decays of HNLs, but no observables can probe the value of $\eta_B$ in this epoch.

Another ingredient that affects dynamics of nuclear reactions is the time-temperature relation, traditionally encoded in the value of $N_\eff$. 
If HNLs have $\tn\ll 0.02$~s, neutrinos are in equilibrium and therefore HNL decays do not change  $N_\eff$, see detailed analysis in~\cite{Boyarsky:2021yoh}.

As a result, evolution of primordial plasma below $T_0^{\text{min}}$ is governed by the SBBN equations.

\section{Dissociation of light elements by mesons and domain of validity of our treatment}
\label{app:mesons-dissociation}
The analysis presented in this paper is valid as long as HNLs decay before neutrino decoupling times. 
This puts the upper limit on HNL lifetime, Eq.~\eqref{eq:hnl-constraint-total}.
It is clear, however, that once the lifetime is so long enough and HNLs (their decay products) can survive till the onset of nuclear reactions, our treatment needs to be changed.

Indeed, pions, if present in the plasma at temperatures $T \lesssim T_{\bbn}$, dissociate light nuclei. The \He threshold-less dissociation processes are (see~\cite{Pospelov:2010cw})
\begin{equation}
 \pi^{-}+\He \to T + n, \quad \pi^{-}+\He \to D + 2n, \quad \pi^{-}+\He \to p + 3n
 \label{eq:mesons-dissociation-processes}
\end{equation}
To estimate estimate the lifetimes at which the processes~\eqref{eq:mesons-dissociation-processes} can be neglected,
we compare the number density of mesons \emph{available for the dissociation} with the number density of \He nuclei:
\begin{equation}
    n^{h}_{\mathrm{He}\text{ diss}}(T_{\bbn})\ll n_{\mathrm{He}}(T_{\bbn}),
    \label{eq:mesons-bottleneck}
\end{equation}
c.f. Eq.~\eqref{eq:mesons-constraint-temp}. Here, $n^{\pi}_{\mathrm{He}\text{ diss}}$ is defined via
\begin{equation}
n^{\pi}_{\mathrm{He}\text{ diss}}(T_{\bbn}) = n_{N}\cdot \Br_{N\to \pi^{-}}\cdot P_{\mathrm{He}\text{ diss}}
\label{eq:npi-inst}
\end{equation}
and $P_{\He\text{ diss}}$ is the probability for a single meson to dissociate \He nuclei before decaying:
\begin{equation}
    P_{\mathrm{He}\text{ diss}} = \frac{\langle\sigma_{\mathrm{He}\text{ diss}}^{\pi}v\rangle n_{\mathrm{He}}}{\Gamma^{\pi}_{\decay}} \simeq 8.3 \cdot 10^{-2} \cdot \frac{ 4 \cdot n_{\mathrm{He}}}{n_B} \left(\frac{T}{\text{1 MeV}}\right)^3,
    \label{eq:p-diss}
\end{equation}
where we used the total cross-section of the dissociation processes~\eqref{eq:mesons-dissociation-processes}, $\langle\sigma_{\mathrm{He}\text{ diss}}^{\pi}v\rangle \simeq 6.5 \cdot F_{\mathrm{He} \pi^-} \text{ mb}$ (a factor $F_{\mathrm{He} \pi^-} \simeq 3.5$ accounts for the Coulomb attraction, Eq.~\eqref{eq:sommerfeld-enhancement}).

In our estimates, we use $T_{\bbn} = 84$ keV, assuming that all free nucleons become bounded in \He nuclei at this temperature. We also do not take into account that after the dissociation of Helium the abundance of lighter elements will be also increased significantly. Assuming $n_{\mathrm{He}}\simeq n_{B}/4$ in~\eqref{eq:mesons-bottleneck}, and using Eqs.~\eqref{eq:npi-inst},~\eqref{eq:p-diss} with the HNL number density given by Eq.~\eqref{eq:number-density-exponential-tail}, we arrive at the upper bound on HNL lifetimes for which our analysis is applicable, $\tau_{N}\lesssim 40\s$.
Notice, that the dependence on $\zeta$ is only logarithmic, therefore its exact value and its mass dependence plays no role.
The analysis for $\tn>40\s$ should be performed separately, see e.g.\ \cite{Bondarenko:2021cpc}.

\end{document}